\UseRawInputEncoding

\documentclass{nature}

\usepackage{comment}

\usepackage{amsmath}
\usepackage[dvipsnames]{xcolor}
\usepackage{soul}
\usepackage{graphicx}
\usepackage{amssymb}
\usepackage{pifont}
\usepackage{longtable}
\usepackage{multirow}
\usepackage{arydshln}
\usepackage{hyperref}
\usepackage{caption}
\usepackage[super,comma,sort&compress]{natbib}
\usepackage{url}

\usepackage[switch, modulo]{lineno}
\modulolinenumbers[1]

\title{The 21-cm forest as a simultaneous probe of dark matter and cosmic heating history}

\author{Yue Shao$^{1}$,
Yidong Xu$^{2,3,\star}$,
Yougang Wang$^{2,3}$,
Wenxiu Yang$^{2,4}$,
Ran Li$^{2,4,5}$,
Xin Zhang$^{1,6,7,\star}$,
Xuelei Chen$^{1,2,3,4,8,\star}$
}

\begin{document}

\maketitle

\begin{affiliations}
 \item Key Laboratory of Cosmology and Astrophysics (Liaoning) \& College of Sciences, Northeastern University, Shenyang 110819, China
 \item National Astronomical Observatories, Chinese Academy of Sciences, Beijing 100101, China
\item Key Laboratory of Radio Astronomy and Technology, Chinese Academy of Sciences, A20 Datun Road, Chaoyang District, Beijing 100101, China
 \item School of Astronomy and Space Science, University of Chinese Academy of Sciences, Beĳing 100049, China
 \item Institute for Frontiers in Astronomy and Astrophysics, Beijing Normal University,  Beijing 102206, China
 \item National Frontiers Science Center for Industrial Intelligence and Systems Optimization, Northeastern University, Shenyang 110819, China
 \item Key Laboratory of Data Analytics and Optimization for Smart Industry (Ministry of Education), Northeastern University, Shenyang 110819, China
 \item Center for High Energy Physics, Peking University, Beijing 100871, China
\end{affiliations}

\begin{abstract}
The absorption features in spectra of high-redshift background radio sources, caused by hyperfine structure lines of hydrogen atoms in the intervening structures, are known collectively as the 21-cm forest. They provide a unique probe of small-scale structures during the epoch of reionization, and can be used to constrain the properties of the dark matter (DM) thought to govern small-scale structure formation. However, the signals are easily suppressed by heating processes that are degenerate with a warm DM model. Here we propose a probe of both the DM particle mass and the heating history of the Universe, using the one-dimensional power spectrum of the 21-cm forest. The one-dimensional power spectrum measurement not only breaks the DM model degeneracy but also increases the sensitivity, making the probe actually feasible. Making 21-cm forest observations with the upcoming Square Kilometre Array has the potential to simultaneously determine both the DM particle mass and the heating level in the early Universe, shedding light on the nature of DM and the first galaxies.
\end{abstract}

\newpage


The 21-cm line of neutral hydrogen (HI) traces various structures throughout cosmic history.
Complementary to the 21-cm tomographic observation, the 21-cm absorption signal against high-redshift
radio point sources probes intervening structures along individual lines of sight \citep{Carilli2002,Furlanetto2002,Furlanetto2006forest,Xu2009,Xu2010,Xu2011MN,Ciardi2013}.
The structures located at different distances along the sightline resembles forest structure on the background source spectrum, which is called 21-cm forest in analogy to the Lyman $\alpha$ (Ly$\alpha$) forest.
The high frequency resolution of radio telescopes allows the 21-cm forest to be a promising probe to small-scale structures during the epoch of reionization (EoR) \citep{Furlanetto2006forest,Xu2011MN}.

In warm dark matter (WDM) models, the small-scale power is suppressed by free-streaming effect compared with the standard cold dark matter (CDM) model \citep{Avila2001,Smith2011,Schneider2013}.
Using Ly$\alpha$ forest as a tracer of small-scale structures, this effect has been used to constrain the WDM particle mass
at low redshifts \citep{Viel2013,Baur2016, Irsic2017}.
Similarly, the 21-cm forest
can potentially be used
deep into the EoR \citep{Shimabukuro2014,Shimabukuro2020},
as the decreased number of low-mass halos leads to weaker 21-cm forest signals.

Methods have been developed to improve the detection of 21-cm forest signal \citep{Mack2012,Ewall2014,Thyagarajan2020}.
However, the 21-cm forest signal can also be suppressed by heating effects during the early galaxy formation\citep{Xu2011MN,Mack2012}.
While this means that it is a sensitive probe of the temperature of the intergalactic medium (IGM)
\citep{Xu2009,Xu2011MN}, it is degenerate with the WDM suppression effect \citep{Shimabukuro2014}, making the interpretation of observations ambiguous.
Nevertheless,
the WDM reduces mainly the number density of 21-cm absorption lines \citep{Shimabukuro2014}, whereas a higher IGM temperature suppresses both the absorption depth and line number density
\citep{Xu2011MN}. This difference makes it possible to distinguish these two effects statistically.

In this Article, we simulate 21-cm forest signals during the EoR under the influence of different dark matter (DM) particle masses and different heating histories of the IGM. We show that although the IGM heating and WDM both suppress the 21-cm signal, they behave differently. By measuring the one-dimensional (1D) power spectrum along lines of sight, it is possible to break the degeneracy, and
constrain the DM particle mass and the IGM temperature (hence the early heating history)
simultaneously.

To simulate the 21-cm forest from the EoR, a high dynamic range is required to model the large-scale structures in density and ionization fields on $\gtrsim 100$ comoving-megaparsec scales, while resolving small-scale halos and their ambient gas on approximately kiloparsec scales.
We use a hybrid approach to achieve this.
The cosmological evolution of large-scale density and ionization fields is simulated with the semi-numerical simulation {\tt 21cmFAST} \citep{Mesinger2011}, with a comoving box size of 1 Gpc with 500$^3$ grids, where the initial density fluctuations are set by DM properties, while each of the $(2~ \mathrm{Mpc})^3$ grid is further divided into 500$^3$ voxels, and populated with halos of various masses according to the local grid density and the conditional halo mass function \citep{Cooray2002,Zentner2007},
which depends on the matter power spectrum regulated by the DM particle mass. The density in each voxel is determined by the Navarro-Frenk-White profile \citep{Navarro1997} or the infall model profile \citep{Barkana2004Infall} according to its distance to the nearest halo (Methods).

\section*{Results}
\label{sec:results}






Recent astrophysical observations have put lower limits on the
WDM particle mass ($m_{\rm WDM}$) of
a few kiloelectronvolts \citep{Viel2013,Menci2017,Garzilli:2019qki,Palanque2020,DES2021}.
We simulate the 21-cm forest signals assuming $m_{\rm WDM} = $ 10 keV, 6 keV, and 3 keV, respectively, to be compared with the signals from a CDM model.
The 21-cm optical depth depends on the density, the neutral fraction of hydrogen gas, and the spin temperature $T_{\rm S}$.
The density field
and ionization field
are simulated according to the DM properties as described above, with more details given in Methods. $T_{\rm S}$ is assumed to be fully-coupled to the gas kinetic temperature $T_{\rm K}$ by the early Ly$\alpha$ background\citep{Wouthuysen1952,Field1959b}, and $T_{\rm K}$ is determined by the heating history of the IGM, or the virial temperature of halos, depending on the gas location (Methods).
The heating history of neutral IGM during the EoR is computed taking into account the adiabatic expansion of the universe, the Compton heating/cooling, and the X-ray heating.
We model the X-ray emissivity as proportional to the formation rate of early non-linear structures \citep{Furlanetto2006global}, normalized by an X-ray production efficiency parameter $f_{\rm X}$
(Methods).

Assuming an unheated IGM ($f_{\rm X} = 0$), the 21-cm optical depth (top panels) and the differential brightness temperature (negative, bottom panels) along a line of sight at $z=9$ are shown in Fig.~\ref{fig_spectrum_WDM}, for CDM (left column) and various WDM particle masses (right columns), respectively.
In the lower panels, the flux density of the background source, scaled to 150 MHz assuming a power-law spectrum, is assumed to be $S_{150}$ = 1 mJy, 10 mJy, and 100 mJy, from top to bottom respectively.
The overall 21-cm absorption depth in WDM models is comparable to the signal level in the CDM model, both corresponding to the absorption depth by the unheated IGM.
However, the small-scale fluctuations are notably reduced in WDM models, due to the more suppressed formation of low-mass halos.
Note that the major contribution to the 21-cm forest signal is from the overdense gas in the halo surroundings which is not heated by virialization shocks \citep{Xu2011MN}.
These small-scale fluctuations are also suppressed,
resulting in sparser absorption lines in the spectra.

Figure~\ref{fig_spectrum_fX} shows the 21-cm optical depth (top panels) and brightness temperature (bottom panels) spectra at $z\sim 9$ in the CDM model, assuming different X-ray efficiency parameters.
As $f_{\rm X}$ increases, the IGM is increasingly heated, increasing the spin temperature and notabley reducing the 21-cm forest signal. The dotted and dashed lines in the lower panels correspond to the thermal noise levels expected for phase-one and phase-two low-frequency
arrays of the Square Kilometre Array (denoted by SKA1-LOW and SKA2-LOW), for which array sensitivities of $A_{\rm eff} / T_{\rm sys} = 800\, {\rm m}^{2} {\rm K}^{-1}$ \citep{deLeraAcedo2020} and $4000\, {\rm m}^{2} {\rm K}^{-1}$ \cite{SKAO}
(with $A_{\rm eff}$ being the total effective area and $T_{\rm sys}$ being the system temperature) are adopted, respectively. For both arrays, we assume a maximum baseline of 65 km, a channel width of 1 kHz, and an integration time of 100 hours (hr).
For the case with negligible early X-rays, the 21-cm forest signal can be marginally detected by the SKA1-LOW for sources with $S_{150} \sim $ 1 mJy, while the same signal will be
easily detected with SKA2-LOW.

However, the heating will notably diminish the detectability of individual absorption lines, weakening the probing power of the 21-cm forest on either the DM properties, or the thermal history of the IGM.
Even if $f_{\rm X}$ = 0.1, i.e. the early star formation has only $\sim 10\%$ X-ray productivity as that of nearby starburst galaxies, the IGM will be heated to about 56 K at $z = 9$, then direct measurement of the 21-cm forest would only be possible for extremely bright quasars
with $S_{150}\gtrsim 100$ mJy for SKA1-LOW, or $S_{150}\gtrsim 10$ mJy for SKA2-LOW, otherwise a much longer integration time would be required.
If $f_{\rm X}\gtrsim 1$, the IGM would be heated to $\gtrsim 650$ K at $z = 9$, then direct detection of the forest signal will
be challenging even for SKA2-LOW.
The heating would be weaker at higher redshifts, but then it would be more difficult to find a suitable quasar as background source.

Moreover, only the fluctuating part of the absorption is measurable in 21-cm forest observation, while the overall absorption depth from the homogeneous IGM would be effectively subtracted when comparing with the intrinsic continuum \citep{Xu2011MN,Ciardi2013}.
If we simply count the absorption lines with a certain threshold of optical depth or equivalent width, the effects of a WDM model and a more heated IGM would be degenerate, both reducing the number of detectable absorbers \citep{Shimabukuro2014}. A statistical variable with more distinguishing power is needed. As we shall show below, the 1D power spectrum of 21-cm forest along the line of sight \citep{Thyagarajan2020} can serve this purpose.

The left panel of Fig.~\ref{fig_PS_fw} compares the 1D power spectra of 21-cm forest in the CDM model with different $f_{\rm X}$.
The 21-cm optical depth is inversely proportional to the gas temperature, and proportional to the density.
As $f_{\rm X}$ increases, the IGM is increasingly heated, and the 1D power spectrum is notably suppressed on all scales.
When the IGM is cold,
the high contrast in temperature between gas in halos and gas in the IGM far from halos dominates the large-scale fluctuations in the optical depth, with typical scales corresponding to the clustering scales of halos of various masses.
As $f_{\rm X}$ increases from 0 to 1, the IGM far from halos with the lowest temperature is heated first, suppressing the temperature contrast on scales of halo clustering, which results in the flattening of 1D power spectrum on large scales.
When $f_{\rm X} = 1$, the IGM temperature is about 650 K at $z = 9$, comparable to the virial temperature ($\sim 1000$ K) of the smallest halos holding gas (with mass $M_{\rm min}\sim$ $10^6$ $M_\odot$, Methods), then
the large-scale fluctuations in the temperature
are mostly smoothed, leaving only a flatter power spectrum originated from density fluctuations. The 21-cm forest and its 1D power spectrum are further reduced when $f_{\rm X}$ increases from 1 to 3.
The 1D power spectra all drop off on small scales corresponding to the clustering scale of the smallest halos holding gas, and the cut-off at the small-scale end is set by the spectral resolution assumed.

The right panel of Fig.~\ref{fig_PS_fw} shows the results for different DM properties assuming an un-heated IGM ($f_{\rm X}$ = 0).
The lower $m_{\rm WDM}$ results in
a much lower level of small-scale density fluctuations, thus suppressing the small-scale 21-cm forest power spectrum. Note that with the same thermal history, the overall amplitude of the 1D power spectrum remains similar for different $m_{\rm WDM}$, while the slope
will be steeper for a warmer DM model. This behavior is distinct from the heating effect, which suppresses the 1D power spectrum more dramatically on all scales.

The dotted and dashed lines in Fig.~\ref{fig_PS_fw} indicate the thermal noise in the power spectrum, $P^{\rm N}$, expected for SKA1-LOW and SKA2-LOW respectively,
utilizing 10 background sources.
The error bars include both the thermal noise of SKA2-LOW and the sample variance
(see Methods). As shown in Fig.~\ref{fig_spectrum_fX}, for a background source with $S_{150} = 10$ mJy, direct measurement of 21-cm forest becomes difficult if $f_{\rm X} \gtrsim 0.1$, and almost impossible even for SKA2-LOW if $f_{\rm X} \gtrsim 1$.
However, the 1D power spectrum of 21-cm forest can be measured precisely by SKA1-LOW over a broad range of wavenumber $k$ if $f_{\rm X}\sim 0.1$,  and it is still detectable by SKA2-LOW with $S_{150} \sim 10$ mJy sources even if $f_{\rm X}$ = 3 at $z=9$.
This is because the absorption appears as an increased variance and can be measured  statistically from the power spectrum even if individual absorbers are too weak to be detected with notableness \citep{Mack2012}. The 1D power spectrum measurement also allows extraction of the scale-dependent information encoded in the density and temperature fields, in contrast to the flatter thermal noise. So the
observation of the 21-cm forest by 1D power spectrum is not only more feasible, but also has better discriminating power for the effects of IGM heating and the WDM.
Fig.~\ref{fig_PS_S150} shows the 1D power spectra for different $f_{\rm X}$ and $m_{\rm WDM}$ assuming $S_{\rm 150}$ = 1 mJy, 10 mJy, and 100 mJy, respectively. Using 1D power spectrum, with 10 background sources of $S_{\rm 150} \sim$ 1 mJy and a moderate integration time of $\sim$ 100 hr, the 21-cm forest signal will be detectable by SKA2-LOW if $f_{\rm X}\lesssim 0.1$, for all DM particle masses considered here.
For brighter sources with $S_{\rm 150} \gtrsim 10$ mJy, the full shape of 1D power spectrum can be well characterized, and a broader range of possible $f_{\rm X}$ values can be probed.
Therefore the 21-cm forest 1D power spectrum will not only break the degeneracy between the effects of WDM and heating, but also be vital to make the probe feasible in practice.

The Universe may also accommodate both a heated IGM and WDM particles, both regulating the amplitude and shape of the 1D power spectrum of 21-cm forest.
We simulate the 
signals for various combinations of $f_{\rm X}$ and $m_{\rm WDM}$ values, and measure the amplitude $P$ and the slope $\beta = {\rm dlog}P(k)/{\rm dlog}k$ of the 1D power spectra at $k = 40\, {\rm Mpc}^{-1}$.
The top panels of Fig.~\ref{fig_magnitude_slope} show that the amplitude of the 1D power spectra roughly determines $f_{\rm X}$, or the IGM temperature, with a weak degeneracy between a higher $f_{\rm X}$ and a smaller $m_{\rm WDM}$. On the other hand, the slope
in the bottom panels shows a different degeneracy;
a flatter power spectrum indicates a higher $f_{\rm X}$ and/or a larger $m_{\rm WDM}$, while a steeper one
implies a lower $f_{\rm X}$ and/or a smaller $m_{\rm WDM}$.
Therefore, the amplitude and slope of 21-cm forest 1D power spectrum can be diagnostic characters for the DM particle mass and the IGM temperature.
When combined, one can effectively break the degeneracy and determine $f_{\rm X}$ and $m_{\rm WDM}$ simultaneously.

With the 21-cm forest 1D power spectrum measured from 100 neutral patches of 10 comoving megaparsec at $z = 9$,
we use the Fisher matrix formalism to forecast constraints on $m_{\rm WDM}$ and $T_{\rm K}$ as expected for both SKA1-LOW and SKA2-LOW, including the thermal noise and sample variance.
Fig.~\ref{fig_contour} shows that if the IGM was only weakly heated, then
very tight constraints can be put on both $m_{\rm WDM}$ and $T_{\rm K}$, with $\sigma_{m_{\mathrm{WDM}}}$ = 1.3 keV and $\sigma_{T_{\mathrm{K}}}$ = 3.7 K for the fiducial model of $m_{\rm WDM} = 6$ keV and $T_{\rm K} = 60$ K after a total observation time of $\delta t$ = 100 hr on each source using SKA1-LOW, and $\sigma_{m_{\mathrm{WDM}}}$ = 0.3 keV and $\sigma_{T_{\mathrm{K}}}$ = 0.6 K using SKA2-LOW. $\sigma_{m_{\rm WDM}}$ and $\sigma_{T_{\rm K}}$ are marginalized absolute errors.
If the IGM was heated up to 600 K at $z = 9$ (corresponding to $f_{\rm X} = 1$), then SKA2-LOW would be required, and we expect to have $\sigma_{m_{\mathrm{WDM}}}$ = 0.6 keV and $\sigma_{T_{\mathrm{K}}}$ = 88 K.
The probe is more sensitive for lower values of $m_{\rm WDM}$.
Note that these constrains can be obtained by measurements on segments of neutral patches along sightlines against 10 background sources with $S_{\rm 150}$ = 10 mJy.
The constraints would be better if more sources
at different redshifts, or brighter sources,
are available.

\section*{Discussion}
\label{sec:discussion}

The 21-cm signal from the EoR can potentially be used to constrain DM properties \cite{Sitwell2014,Shimabukuro2014,Munoz2020,Hibbard2022},
but the degeneracies with astrophysical effects can be an obstacle\cite{Sitwell2014,Shimabukuro2014,Munoz2020}.
During the EoR, there are various feedback effects \cite{CFreview2005}. Here we consider primarily radiative feedbacks, including Ly$\alpha$ photons coupling $T_{\rm S}$ to $T_{\rm K}$, ionizing photons determining the large-scale ionization field, and X-ray photons heating the IGM. The mechanical and chemical feedbacks affect the density profiles and the cooling mechanisms, but have minor influences on the 21-cm forest.
The main focus of this work is the heating effect that is most important in reducing the 21-cm forest signal and is degenerate with the WDM effect.

Using a set of semi-numerical simulations covering a high dynamic range,
we show that both the presence of WDM and an early X-ray heating can reduce the number of observable 21-cm absorbers. This degeneracy hinders the 21-cm forest from being an effective probe to either the DM properties or the thermal history of the universe.
We have demonstrated that the 1D power spectrum of 21-cm forest is a good observable to break this degeneracy, and is even effective in high heating-rate cases in which the number of 21-cm forest lines is severely diminished.
By quantifying the fluctuations, the 1D power spectrum of the 21-cm forest is also immune to subtraction of the overall absorption from the homogeneous IGM in practical observations. The DM particle mass and the IGM temperature at a specific redshift can be simultaneously constrained.

Although in our simulation the gas density profile surrounding a halo is based on simple models,
this does not have much impact on the number density and the clustering properties of absorption lines, which determines the main characteristics of the 1D power spectrum.
We also note that the overall signal level is dependent on the local density $\delta_0$ in the large-scale environment.
We investigate the effect of local density by computing the 21-cm forest signals on different grids, with various densities on the 2 Mpc scale.
As shown in Extended Data Figs.~1 and 2, the local density affects the overall magnitude of signals,
but the effect is much weaker than the heating, even in the extreme case of $\delta_0 = 2$ in a grid of $\sim 2$ Mpc.
Meanwhile, the local density has almost negligible effect on the shape of 1D power spectrum, making the effect distinguishable from the WDM effect.

While direct detection of individual 21-cm absorption lines will be challenging if the early IGM is heated, the 1D power spectrum measurement is more promising. The observation relies on the availability of high-redshift radio-bright sources prior to reionization.
Quite a number of radio-loud quasars have been detected beyond redshift 5 \citep{Shao2020GMRT,Liu2021RLF,Shao2022QSO}, including nine at $z > 6$\citep{QSOz682,QSOz644,Gloudemans2022}.
A few hundred radio quasars with $> 8$ mJy at $z \sim 6$ are expected to be spectroscopically observed in the near future \cite{Gloudemans2021}. As there is no evidence for the evolution in the radio loudness fraction of high-$z$ quasars \citep{Liu2021RLF,Gloudemans2021}, one can expect about $\sim 2000$ sources
with $> 6$ mJy at $8 < z < 12$ \citep{Haiman2004}.
The long-duration gamma-ray bursts (GRBs)
are also possible high-redshift sources.
Several cases have been discovered beyond redshift 8 \citep{GRB090423z8,GRB090429Bz9}.
For future missions like the High-z Gamma-ray bursts for Unraveling the Dark Ages Mission and the Transient High-Energy Sky and Early Universe Surveyor, the expected detection rate of luminous GRBs from Population III stars is 3 – 20  $\rm{yr}^{-1}$ at $z > 8$
\citep{GRBhighz2019}.
Given the higher sensitivity of  1D power spectrum observation, radio afterglows of high-$z$ GRBs could also be used.
The fast radio bursts, though brighter, are however too brief to allow long integration required.

Current combination of astrophysical probes of strong gravitational lensing, Ly$\alpha$ forest, and luminous satellites of our Galaxy
indicates that $m_{\rm WDM}$ may be larger than 6 keV\citep{Enzi2021}, but models with $m_{\rm WDM}$ of a few keV are still not excluded.
On the other hand,
tomographic 21-cm power spectrum measurement, in combination with complementary probes, yield a constraint on the IGM temperature of $8.9 \,{\rm K} < T_{\rm K} < 1.3\times 10^3 \,{\rm K}$ at $z\sim 8$ at 68\% confidence\citep{HERAconstrain2022}.
With the upcoming SKA-LOW,
the 21-cm forest observation, especially the 1D power spectrum, can improve the constraints on
both the properties of DM and the thermal history of the early universe simultaneously, providing an effective probe to the DM in an unexplored era in the structure formation history,
and to the first galaxies interplaying with the early IGM.

\clearpage





\clearpage

\section*{Methods}

\subsection{The 21-cm forest signal.}
\label{sub:21cmsig}

Using high-redshift quasars or radio afterglows of GRBs as background radio sources \citep{Carilli2002,Xu2009}, the HI in halos and in the IGM absorbs 21-cm photons along the line of sight.
The 21-cm forest signal is the flux decrements due to 21-cm absorption with respect to the continuum of a background radio source, which in the Rayleigh-Jeans limit is characterized by the differential brightness temperature.
In the optically-thin limit, which is usually the case for the 21-cm transition, the observed differential brightness of the 21-cm absorption signal, relative to the brightness temperature of the background radiation $T_{\gamma}(\hat{\boldsymbol s}, \nu_0, z)$ at a specific direction $\hat{\boldsymbol s}$ and redshift $z$, is
\begin{align}\label{eq.Tb}
\delta T_{\rm b}(\hat{\boldsymbol{s}}, \nu) \approx \frac{T_{\rm S}(\hat{\boldsymbol{s}}, z)-T_{\rm \gamma}(\hat{\boldsymbol{s}}, \nu_0, z)}{1+z} \tau_{\nu_0}(\hat{\boldsymbol{s}}, z).
\end{align}
Here $\nu_0$ = 1420.4 MHz is the rest-frame frequency of 21-cm photons, $T_{\rm S}$ is the spin temperature of the absorbing HI gas, and $\tau_{\nu_0}$ is the 21-cm optical depth.
In terms of the average gas properties within each voxel, the 21-cm optical depth can be written as \citep{Field1959a,Madau1997,Furlanetto2006forest}
\begin{align}\label{eq.tau}
\tau_{\nu_0}(\hat{\boldsymbol{s}}, z) \approx
0.0085\left[1+\delta(\hat{\boldsymbol{s}}, z)\right] (1+z)^{3/2}
\left[\frac{x_{\rm HI}(\hat{\boldsymbol{s}}, z)}{T_{\rm S}(\hat{\boldsymbol{s}}, z)}\right] \left[\frac{H(z) /(1+z)}{{\rm d} v_{\|} / {\rm d} r_{\|}}\right]
\left(\frac{\Omega_{\rm b}h^2}{0.022}\right)\left(\frac{0.14}{\Omega_{\rm m}h^2}\right),
\end{align}
where $\delta(\hat{\boldsymbol{s}}, z)$, $x_{\rm HI}(\hat{\boldsymbol{s}}, z)$, and $H(z)$ are the gas overdensity, the neutral fraction of hydrogen gas, and the Hubble parameter, respectively, and ${\rm d} v_{\|}/{\rm d} r_{\|}$ is the gradient of the proper velocity projected to the line of sight. $\Omega_{\rm b}$, $\Omega_{m}$ and $h$ are baryon density parameter, matter density parameter and dimensionless Hubble constant, respectively.

The brightness temperature of the background radiation at the rest frame of the 21-cm absorption $T_{\gamma}(\hat{\boldsymbol{s}}, \nu_0, z)$ is related to the observed brightness temperature at a redshifted frequency $\nu$,  $T_{\gamma}(\hat{\boldsymbol{s}}, \nu, z=0)$, by $T_{\gamma}(\hat{\boldsymbol{s}}, \nu_0, z)=(1+z) T_{\gamma}(\hat{\boldsymbol{s}}, \nu, z=0)$, and it has contributions from both the background point source and the cosmic microwave background (CMB), i.e.
$T_{\gamma}(\hat{\boldsymbol{s}}, \nu, z=0)=T_{\rm rad}(\hat{\boldsymbol{s}}, \nu, z=0)+T_{\rm CMB}(z=0)$,
where $T_{\rm rad}(\hat{\boldsymbol{s}}, \nu, z=0)$ represents the observed brightness temperature of the point source, and it usually dominates over the CMB temperature ($T_{\rm CMB}$).

For a given radio telescope resolving a solid angle of $\Omega$,
the observed brightness temperature of a source is related to the flux density $S_{\rm rad}(\nu)$ by
\begin{align}
T_{\rm rad}(\hat{\boldsymbol{s}}, \nu, z=0) =\frac{c^2}{2 k_{\rm B} \nu^{2}}\frac{S_{\rm rad}(\nu)}{ \Omega},
\end{align}
where $c$ is the speed of light and $k_{\rm B}$ is the Boltzmann constant.
The flux density of the background source is modeled to have a power-law spectrum scaled to 150 MHz, i.e. $S_{\rm rad}(\nu)$ = $S_{150}\left(\nu / \nu_{150}\right)^{\eta}$ \citep{Thyagarajan2020}, where $\nu_{150}$ = 150 MHz and a spectral index of $\eta=-1.05$ is assumed as appropriate for a powerful radio source like Cygnus A  \citep{Carilli2002}.
Note that the spectral index of high-redshift quasars has a large scatter, and their spectra may be flatter than Cygnus A at low frequencies \citep{Shao2020GMRT,QSOz644}, but the detailed spectral index makes only a negligible difference to our results.
In this work, we take the flux densities of $S_{150} =$ 1 mJy, 10 mJy, and 100 mJy for the background point sources as examples, and assume the maximum baseline of 65 km for both the SKA1-LOW and SKA2-LOW for calculating the angular resolution for a given redshift.

Assuming that $T_{\rm S}$ is fully coupled to $T_{\rm K}$ by the early Ly$\alpha$ background, the 21-cm  optical depth $\tau_{\nu_0}$ and the forest signal $\delta T_{\rm b}$ are then dependent on the density
$\delta$, neutral fraction $x_{\rm HI}$, gas temperature $T_{\rm K}$, and the velocity gradient ${\rm d} v_{\|}/{\rm d} r_{\|}$, of each voxel along the line of sight.
Here we account only for the Hubble expansion for the velocity field, but neglect the peculiar velocity,
as the peculiar velocity mainly shifts the contribution of the absorption from individual segments of gas. We note that the peculiar velocity may affect the individual line profiles \citep{Xu2011MN}, but we expect that its effect on the overall amplitude of the signal and the 1D power spectrum is small. The density field, ionization field, and the gas temperature field are modeled as follows.
Throughout this study, we adopted the set of cosmological parameters consistent with the Planck 2018 results\citep{Planck2018}: $\Omega_{\rm m}$ = 0.3153, $\Omega_{\rm b} h^2$ = 0.02236, $\Omega_{\Lambda}$ = 0.6847, $h$ = 0.6736, $\sigma_8$ = 0.8111. $\Omega_{\rm \Lambda}$ and $\sigma_8$ are dark-energy density parameter and matter fluctuation amplitude, respectively.

\subsection{The density field.}
\label{sub:density}
The evolution of the large-scale density field is simulated with linear theory using the
{\tt 21cmFAST} \citep{Mesinger2011}, for both the CDM and WDM models.
The simulation box has a comoving size of ${\rm (1 Gpc)^3}$, and ${\rm (500)^3}$ grids.
The influence of DM properties on the density field is mainly on small scales.
In each of the 2 Mpc grids, the small-scale density distribution is simulated by randomly populating halos according to the conditional halo mass function and the local density of the grid from the {\tt 21cmFAST} simulation, and assigning density profiles to the gas in the halos as well as in the IGM, as detailed below.

\subsubsection{Halo mass function.}
In the framework of the CDM model, the number density of halos per mass interval in the range $(M, M + {\rm d}M)$, in a simulation grid with mass $M_0$ and overdensity $\delta_0$ at redshift $z$, can be modeled by the conditional halo mass function \citep{Cooray2002,Zentner2007} of the Press-Schechter form \citep{Press1974}, i.e.
\begin{align}\label{eq.CMF_CDM}
\frac{{\rm d} n(M|\delta_0,M_0;z)}{{\rm d} M}=\sqrt{\frac{1}{2 \pi}} \frac{\bar{\rho}_{\rm m0} (1+\delta_0)}{M} \left|\frac{{\rm d} S}{{\rm d} M}\right|
\frac{\delta_{\rm c}(z) - \delta_0}{(S-S_0)^{3/2}}\, \exp{\left\{-\frac{[\delta_{\rm c}(z) - \delta_0]^2}{2(S-S_0)}\right\}},
\end{align}
where $\bar{\rho}_{\rm m0}$ is the average density of matter in the universe today, $S=\sigma^2(M)$ is the variance of mass scale $M$, $S_0=\sigma^2(M_0)$, and $\delta_{\rm c}(z)=1.686/D(z)$ is the critical overdensity for collapse at redshift $z$ extrapolated to the present time using the linear theory, in which $D(z)$ is the linear growth factor.

In the WDM model, the structure formation is suppressed below the free streaming scale $\lambda_{\rm fs}$ of DM particles, and the conditional halo mass function can be approximately written as \citep{Smith2011}
\begin{align}\label{eq.CMF_WDM}
\frac{{\rm d} n(M|\delta_0,M_0;z)}{{\rm d} M}=\frac{1}{2}\left\{1+\operatorname{erf}\left[\frac{\log _{10}\left(M / M_{\rm fs}\right)}{\sigma_{\log M}}\right]\right\}\left[\frac{{\rm d} n (M|\delta_0,M_0;z)}{{\rm d} M}\right]_{\rm PS},
\end{align}
where $\sigma_{\log M}=0.5$, and
$M_{\rm fs}$ is the suppressing mass scale of halo formation corresponding to $\lambda_{\rm fs}$, i.e.
$M_{\rm fs}=4 \pi (\lambda_{\rm fs}/2)^{3} \rho_{\rm m0} /3$. PS represents Press-Schechter form in CDM model.
The comoving free streaming scale is approximately \citep{Smith2011}
\begin{align}
\lambda_{\rm fs} \approx 0.11\left(\frac{\Omega_{\rm WDM} h^{2}}{0.15}\right)^{1 / 3}\left(\frac{m_{\rm WDM}}{\rm keV}\right)^{-4 / 3}({\rm Mpc}),
\end{align}
where $\Omega_{\rm WDM}$ is the WDM density normalized by the critical density.
The Press-Schechter mass function $[{\rm d} n (M|\delta_0,M_0;z)/{\rm d} M]_{\rm PS}$ in Eq.~(\ref{eq.CMF_WDM}) takes the form of Eq.~(\ref{eq.CMF_CDM}), but the variance of density fluctuations is evaluated with the  matter power spectrum fitted for WDM \citep{Bode2001}:
\begin{align}
P_{\rm WDM}(k)=P_{\rm CDM}(k)\left\{\left[1+(\alpha k)^{2 \beta}\right]^{-5 / \beta}\right\}^{2},
\end{align}
where $\beta$ = 1.12 and $\alpha$ is given by \citep{Viel2005}
\begin{align}
\alpha=0.049\left(\frac{m_{\rm WDM}}{\rm keV}\right)^{-1.11} \left(\frac{\Omega_{\rm WDM}}{0.25}\right)^{0.11}\left(\frac{h}{0.7}\right)^{1.22} h^{-1}\, ({\rm Mpc}).
\end{align}

Supplementary Fig.~1 shows the halo mass function, evaluated at $\delta_0 = 0$ and $S_0 = 0$, for both CDM and WDM models.
The halo number is obviously suppressed below the free streaming scale in the WDM models, with the lower $m_{\rm WDM}$ resulting in larger suppressing scale.
Especially, the WDM models notably reduce the total number of halos by suppressing the small ones, thus suppressing the small-scale fluctuations in the neutral hydrogen density, which have a major contribution to the 21-cm forest signals.

The major contribution to the 21-cm forest signal comes from the gas  in and around the large number of low-mass halos that are not producing ionizing photons and reside in neutral environments\citep{Xu2011MN,Shimabukuro2014}.
Therefore,
we focus on neutral patches along a given line of sight, and select neutral grids from the large-scale ionization field simulated by {\tt 21cmFAST}. Then we randomly populate each of these 2 Mpc grids with halos according to the conditional mass function determined by the DM models.
We consider only the halos with
the mass upper limit $M_4$ corresponding to the virial temperature of $T_{\rm vir}$ = $10^4$ K, so that the atomic cooling is not efficient enough to enable substantial star formation. The lower limit of halo mass $M_{\rm min}$ is set by the filtering mass scale, so that the halos could retain most of its gas and the gas in the ambient IGM to contribute to the 21-cm absorption.
The filtering mass is mainly determined by thermal history of the universe, and it is of order $\sim 10^6 M_{\odot}$ for the redshifts of interest ($7\lesssim z\lesssim 11$) for $f_{\rm X}\lesssim 1$ in the CDM model\citep{Xu2011MN}. It would be higher for higher $f_{\rm X}$, and the different density profiles in WDM models may also slightly modify its value. In the present work, we set the same $M_{\rm min} = 10^6 M_{\odot}$ for all the models for simplicity, but we expect that the dependence of filtering mass on $f_{\rm X}$ will make the probe more sensitive to the thermal history of the universe, while more challenging to discriminate WDM models for cases with high $f_{\rm X}$.

\subsubsection{Gas profile.}
Each grid along the line of sight is further divided into ${\rm (500)^3}$ voxels, each with a size of ${\rm (4\,kpc)^3}$, then the gas density of each voxel is determined by its distance to the nearby halos.
Inside the virial radius $r_{\rm vir}$, we assume that the dark matter follows the NFW density profile \citep{Navarro1997},
and the gas is in hydrostatic equilibrium with the dark matter \citep{Xu2011MN}.
Thus, the gas density distribution can be derived analytically \citep{Makino1998}:
\begin{align}
\ln \rho_{\rm g}(r)=\ln \rho_{\rm gc}-\frac{\mu m_{\rm p}}{2 k_{\rm B} T_{\rm vir}}\left[v_{\rm e}^{2}(0)-v_{\rm e}^{2}(r)\right],
\end{align}
where $\rho_{\rm gc}$ denotes the central gas density, $\mu$ is the mean molecular weight of the gas, $m_{\rm p}$ is the proton mass,
and $v_{\rm e}(r)$ is the gas escape velocity at radius $r$,
given by
\begin{align}
v_{\rm e}^{2}(r)=2 \int_{r}^{\infty} \frac{G M\left(r^{\prime}\right)}{r^{\prime 2}} {\rm d} r^{\prime}=2 V_{\rm c}^{2} \frac{F(y x)+\frac{y x}{1+y x}}{x F(y)}.
\end{align}
Here $V_{\rm c}^2 \equiv G M/r_{\rm vir}$ is the circular velocity at the virial radius, $G$ is gravitational constant,
$x \equiv r / r_{\rm vir}$,  $y$ is the halo concentration, and $F(y)=\ln(1+y)-y/(1+y)$.
The central gas density is determined by normalizing the total baryonic mass fraction of the halo to the cosmic mean value, which gives
\begin{align}
\rho_{\rm gc} =\frac{\left(\Delta_{c} / 3\right) {y}^{3}\left(\Omega_{\rm b} / \Omega_{\rm m}\right) e^{A}}{\int_{0}^{y}(1+t)^{A / t} t^{2} {\rm d} t} \bar{\rho}_{\rm m}(z),
\end{align}
where $\bar{\rho}_{\rm m}(z)$ is the mean matter density of the universe at redshift $z$, $A \equiv 2 y/F(y)$, $e$ is the mathematical constant (base of natural log), and $\Delta_{c}=18 \pi^{2} + 82\left(\Omega_{\rm m}^{z}-1\right) - 39\left(\Omega_{\rm m}^{z}-1\right)^{2}$ is the mean density of a virialized halo with respect to the cosmic mean value \citep{Bryan1998}, in which $\Omega_{\rm m}^{z}=\Omega_{\rm m}(1+z)^{3} /\left[\Omega_{\rm m}(1+z)^{3}+\Omega_{\rm \Lambda}\right]$.

The gas density in the halo surroundings is enhanced because of the gravitational potential.
Outside the virial radii of halos, we assume that the gas density profile follows the dark matter distribution,
and it can be computed by using the infall model which is based on the excursion set theory \citep{Barkana2004Infall}. The gas density profiles in and around halos of different masses are plotted in
Supplementary Fig.~2 for $z = 9$.
It is seen that there is density discontinuity at the virial radius in our model. This is expected at the virialization shock near the virial radius \citep{Abel2002,Keshet2003}, though the exact location of the shock may vary from halo to halo \citep{Abel2002}.

The infall model was developed for the matter density and velocity distribution around density peaks \citep{Barkana2004Infall}. Directly applying it to arbitrary environments may over-predict the gas density in under-dense regions. Therefore,
we normalize the density field to ensure that the minimum density is 0, and the average density of the $(500)^3$ voxels in each 2 Mpc grid equals the grid density from the large-scale {\tt 21cmFAST} simulation.
To test the reliability for the small-scale density field, we run a small-scale high-resolution hydrodynamical simulation with the {\tt GADGET} (GAlaxies with Dark matter and Gas intEracT) \citep{Springel2005} for high redshifts. The simulation has a box size of 4 $h^{-1}{\rm Mpc}$ and $2\times800^3$ gas and DM particles \citep{Xu2021}.
We compare the probability density distribution of our analytical gas density field with the one from the simulated gas density
in Supplementary Fig.~3,
at the same resolution at $z$ = 17.
It shows that our gas density model closely recovers the probability distribution of the gas density fluctuations from the hydrodynamical simulations.

The line-of-sight density distribution in the CDM model is illustrated in the left panel of Extended Data Fig.~1
for three grids with different local overdensities $\delta_0$ on the 2 Mpc scale at $z=9$.
The density distributions for different DM properties are shown in
Supplementary Fig.~4.

\subsection{The ionization field.}
\label{sub:ion}
The large-scale ionization field is simulated with the semi-numerical simulation {\tt 21cmFAST} assuming ionizing sources with a minimum halo mass of $M_4$ and an ionizing efficiency parameter of $\zeta = 11$ \citep{Mesinger2011}.
By suppressing the formation of small-scale halos, the WDM models may possibly speed up or delay the large-scale reionization process by modifying both the abundances of ionizing sources and sinks \citep{Yue2012,Dayal2017}.
In the present work, we use the basic version of {\tt 21cmFAST} in which the effect of sinks is incorporated by a homogeneous recombination number, and the reionization is delayed in the WDM models as shown in Supplementary Fig.~5.
It shows that the effect of WDM on the large-scale reionization history becomes obvious only if
$m_{\rm WDM} \lesssim 3$ keV, and this is consistent with the fact that atomic-cooling halos are effectively suppressed in WDM models with $m_{\rm WDM} \lesssim 3$ keV as shown in Supplementary Fig.~1.
Note that
the 21-cm forest signals mainly come from neutral regions,
and we pick up neutral patches in the large-scale simulation box to analyze the small-scale structures in the 21-cm forest signals.
The large-scale reionization history only determines the probability of getting a neutral patch of the IGM with a certain length along a line of sight.
In order to have consistent source properties when comparing the results for the same $f_{\rm X}$, we set the same ionizing efficiency parameter for all the models considered here, while the global reionization history would be slightly different among WDM and CDM models. On the other hand, a different reionization scenario may change the minimum source mass, for example, in a reionization model with stronger feedback effects would have a minimum halo mass for collapse higher than $M_4$, thus changing the reionization history. However,
the large-scale ionization field and the overall reionization history have only a minor effect on the small-scale 21-cm forest signals we are interested in.

For each of the neutral grids in the simulation box, we assume that the gas is in collisional ionization equilibrium (CIE), so that the ionized fraction of each voxel is determined by its local density and temperature, i.e.
\begin{align}
n_{\rm e} n_{\rm HI} \gamma=\alpha_{\rm B} n_{\rm e} n_{\rm p},
\end{align}
where $n_{\rm HI}$, $n_{\rm e}$ and $n_{\rm p}$ represent the number densities of neutral hydrogen, electron and proton, respectively, $\gamma$ is the collisional ionization coefficient \citep{Cen1992}, and $\alpha_{\rm B}$ is the case B recombination coefficient \citep{Hui1997} which is appropriate for low-mass halos and the incompletely ionized IGM.
Here both $\gamma$ and
$\alpha_{\rm B}$ are functions of temperature.

\subsection{The temperature field.}
\label{sub:tem}

The gas temperature $T_{\rm K}$ of each voxel is determined by the thermal history of the early universe and the location of the voxel with respect to halos.
While the photoionization heating by the UV background dominates the gas heating in ionized regions \citep{Hui2003}, it is the X-rays that can penetrate deep into the neutral IGM and dominate the heating of the neutral gas contributing to 21-cm signals.
For the gas in the neutral
IGM, its temperature is mainly determined by the cosmic expansion, the heating or cooling from the Compton scattering, and the X-ray heating.
The global evolution of the IGM temperature can be written as \citep{Furlanetto2006global}
\begin{align}
\frac{{\rm d} T_{\rm K}}{{\rm d} t}=-2 H(z) T_{\rm K}+\frac{2}{3} \frac{\epsilon_{\rm comp}}{k_{\rm B} n}+\frac{2}{3} \frac{\epsilon_{\rm X,h}}{k_{\rm B} n},
\end{align}
where $n$ is the total particle number density,
$\epsilon_{\rm comp}$ is the Compton heating/cooling rate per unit physical volume \citep{Furlanetto2006global}, and  $\epsilon_{\rm X,h}$ represents
the part of the X-ray emissivity $\epsilon_{\rm X}$ that contributes to heating,
for which we adopt a fitted formula to simulations, i.e.
$\epsilon_{\rm X,h} = [1-0.8751(1-{x_i}^{0.4052})] \epsilon_{\rm X}$ \citep{Valdes2008}, where $x_i$ is the ionized fraction.
Assuming that the X-ray productivity is proportional to the star formation rate, and hence to the matter collapse rate, the total X-ray emissivity $\epsilon_{\rm X}$ can be written as \citep{Furlanetto2006global}:
\begin{align}
\frac{2}{3} \frac{\epsilon_{\rm X}} {k_{\rm B} n\, H(z)} = 5 \times 10^{4} {\rm ~K} f_{\rm X} \left(\frac{f_{\star}}{0.1} \frac{{\rm d} f_{\rm coll} / {\rm d} z}{0.01} \frac{1+z}{10}\right).
\end{align}
Here $f_{\star}$ is the star formation efficiency approximately evaluated at $M_4$ \citep{Salvadori2009}, as appropriate for the most abundant star-forming halos, $f_{\rm coll}$ is the fraction of matter  collapsed into atomic-cooling halos with $M>M_4$, and $f_{\rm X}$ is the normalization parameter describing the uncertain nature of X-ray productivity in the early universe as compared to the local universe\citep{Furlanetto2006global,Mack2012}.
The global evolution of the IGM temperature $T_{\rm K}$ is shown in
Supplementary Fig.~6 for different values of $f_{\rm X}$. The curve with $f_{\rm X}$ = 0 denotes the case with purely adiabatic cooling and Compton heating.

Inside the virial radius, the gas kinetic temperature $T_{\rm K}$ equals to the virial temperature $T_{\rm vir}$ of the halo.
As for the gas in the overdense regions near halos, it will be adiabatically heated depending on the local density.
In the absence of X-rays, the temperature profiles for halos with $10^6 M_\odot$, $10^7 M_\odot$, and $10^8 M_\odot$ are illustrated in
Supplementary Fig.~7 for $z = 9$.
Similar to the density profiles, the gas temperature also shows discontinuity at the virialization shocks as expected, but the exact location of the virialization shocks has negligible effects on our main results.
In the cases with X-ray heating, the gas temperature outside the halos is set by the maximum between the adiabatic temperature and the heated IGM temperature.

\subsection{Thermal noise of direct measurement.}
\label{sub:noise}

\begin{figure}
\centering
\includegraphics[angle=0, width=16.0cm]{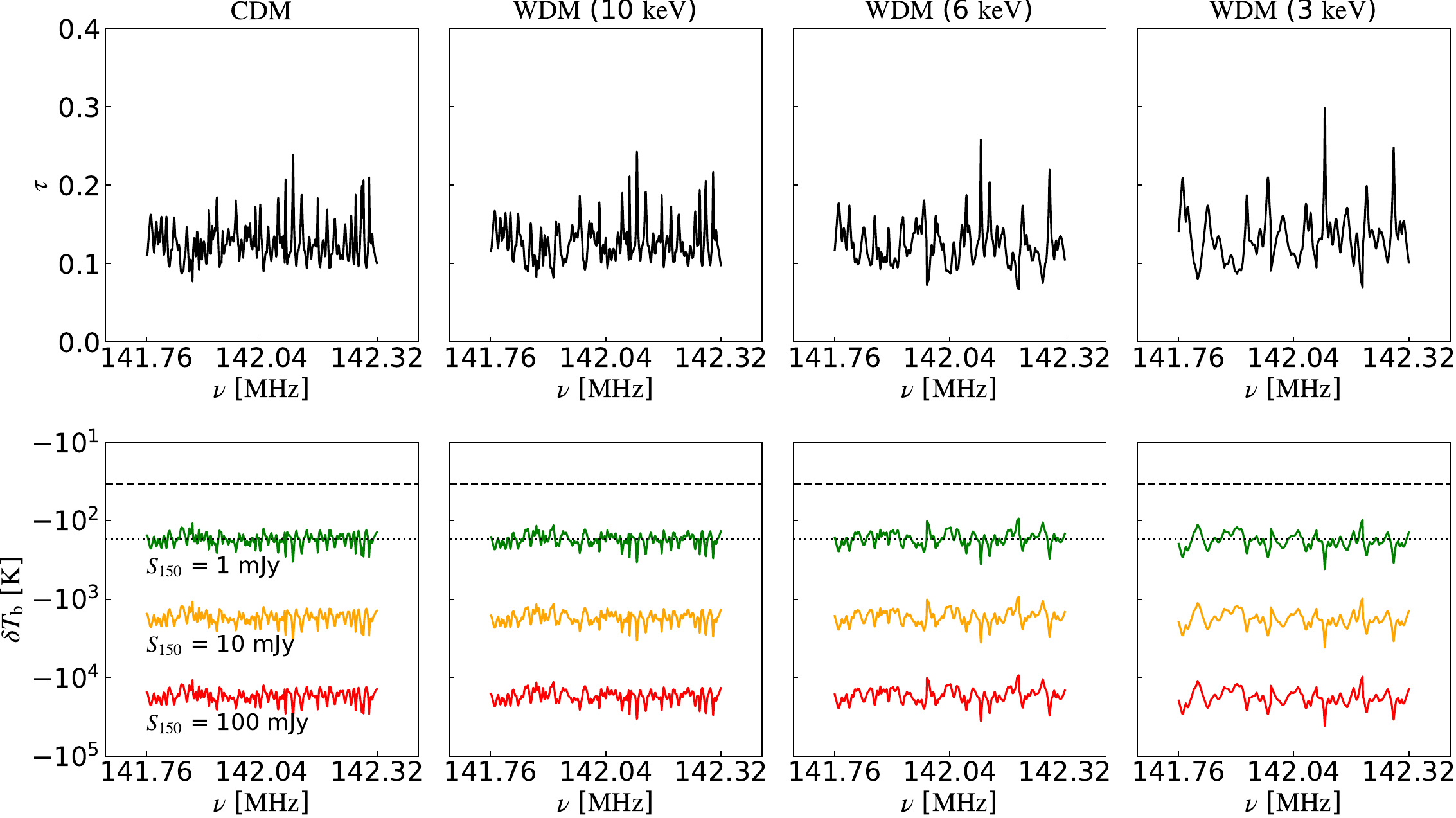}
\caption{\label{fig_spectrum_WDM}
  \textbf{Synthetic spectra for an unheated IGM.}
  The synthetic spectra of optical depth (top) and brightness temperature (bottom) for a neutral patch of 10 comoving megaparsec along the line of sight, for an un-heated IGM ($f_{\rm X} = 0$) at $z$ = 9.
  In each row, the four columns correspond to the CDM model, and the WDM models with $m_{\rm WDM}$ = 10 keV, 6 keV, and 3 keV, from left to right respectively.
  In the lower panels, the green, yellow, and red spectra correspond to the background source flux densities of $S_{150} = $ 1 mJy, 10 mJy, and 100 mJy, respectively.
  The spectra have been smoothed with a channel width of 1 kHz, and
  the dotted and dashed lines are the thermal noise levels $\delta T^{\rm N}$ expected for SKA1-LOW and SKA2-LOW respectively, with an integration time of $\delta t = 100$ hr.
  }
\end{figure}

\begin{figure}
\centering
\includegraphics[angle=0, width=16.0cm]{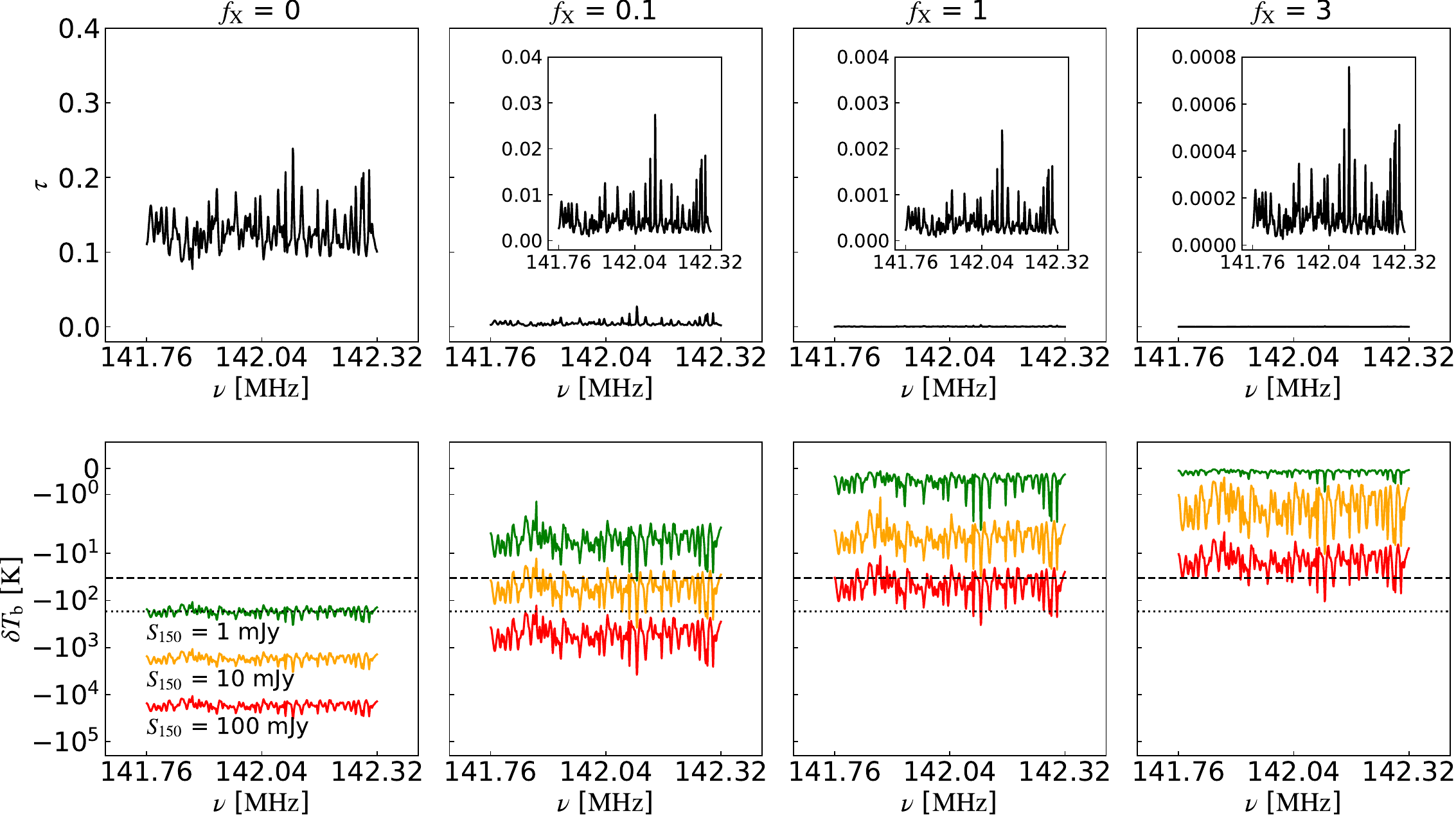}
\caption{\label{fig_spectrum_fX}
  \textbf{Synthetic spectra for the CDM model.}
  The synthetic spectra of optical depth (top) and brightness temperature (bottom) for a neutral patch of 10 comoving megaparsec along the line of sight, for the CDM model at $z$ = 9.
  In each row, the four columns correspond to $f_{\rm X}$ = 0, 0.1, 1, and 3, from left to right respectively.
  In the lower panels, the green, yellow, and red spectra correspond to $S_{150} = $ 1 mJy, 10 mJy, and 100 mJy, respectively,
  and the dotted and dashed lines are the thermal noise levels $\delta T^{\rm N}$ expected for SKA1-LOW and SKA2-LOW respectively, with $\delta \nu = 1$ kHz and $\delta t = 100$ hr.
  The zoom-in plots in the top panels show the 21-cm optical depth with different scales on the y axes.
 }
\end{figure}

In the direct measurement of individual absorption lines,
the noise flux density averaged over two polarizations can be expressed as \citep{Thompson2017}:
\begin{align}
\delta S^{\rm N} \approx \frac{2 k_{\rm B} T_{\rm sys}}{A_{\rm eff} \sqrt{2 \delta \nu \delta t}},
\end{align}
where $A_{\rm eff}$ is the effective collecting area of the telescope, $T_{\rm sys}$ is the system temperature, $\delta \nu$ is the channel width, and $\delta t$ is the integration time.
The corresponding thermal noise temperature is:
\begin{align}
\delta T^{\rm N} = \delta S^{\rm N} \left(\frac{\lambda_{z}^2 }{2 k_{\rm B} \Omega}\right)
\approx \frac{\lambda_{z}^2  T_{\rm sys}}{A_{\rm eff} \Omega \sqrt{2 \delta \nu \delta t}},
\end{align}
where $\lambda_{z}$ is the observed wavelength, and $\Omega=\pi (\theta/2)^2$ is the solid angle of the telescope beam, in which $\theta = 1.22\lambda_{z}/D$ is the angular resolution with $D$ being the longest baseline of the radio telescope/array.
For the SKA1-LOW, we adopt $A_{\rm eff} / T_{\rm sys}= 800\, {\rm m}^{2} {\rm K}^{-1}$ \citep{deLeraAcedo2020},
and $A_{\rm eff} / T_{\rm sys}= 4000\, {\rm m}^{2} {\rm K}^{-1}$ is expected for SKA2-LOW \cite{SKAO}.
For both arrays, we assume $D = 65$ km and $\delta t$ = 100 hr, and $\delta \nu = 1$ kHz is assumed in order to resolve individual 21-cm lines.
Correspondingly, the synthetic spectra shown in Figs.~\ref{fig_spectrum_WDM} and \ref{fig_spectrum_fX} are smoothed with the same channel width.
At redshift $z$ = 9, the angular resolution is about 8.17 arcsec, and the noise temperature is plotted with dotted and dashed lines in the lower panels in
Figs.~\ref{fig_spectrum_WDM} and \ref{fig_spectrum_fX}, for SKA1-LOW and SKA2-LOW respectively.

\subsection{1D power spectrum of 21-cm forest.}
\label{sub:ps}
It is seen from Fig.~\ref{fig_spectrum_fX} that the direct measurement of individual absorption lines is vulnerably hampered by the early X-ray heating. In order to improve the sensitivity for detecting the 21-cm forest signal, and to reveal the clustering properties of the absorption lines so as to distinguish the effects between heating and WDM models, we follow the algorithm in Ref.~\cite{Thyagarajan2020}, and compute the 1D power spectrum of the brightness temperature on hypothetical spectra against high-redshift background sources.
The brightness temperature $\delta T_{\text {b}}(\hat{\boldsymbol{s}}, \nu)$ as a function of observed frequency $\nu$ can be equivalently expressed in terms of line-of-sight distance $r_z$, $\delta T_{\rm b}^{\prime}\left(\hat{\boldsymbol{s}}, r_{z}\right)$, and the Fourier transform of $\delta T_{\text {b }}^{\prime}\left(\hat{\boldsymbol{s}}, r_{z}\right)$ is
\begin{align}
\delta \widetilde{T}^{\prime}\left(\hat{\boldsymbol{s}}, k_{\|}\right)=\int \delta T_{\rm b}^{\prime}\left(\hat{\boldsymbol{s}}, r_{z}\right) e^{-i k_{\|} r_{z}} {\rm ~d} r_{z}.
\end{align}
The 1D power spectrum along the line of sight is defined as:
\begin{align}
P\left(\hat{\boldsymbol{s}}, k_{\|}\right) = \left|\delta \widetilde{T}^{\prime}\left(\hat{\boldsymbol{s}}, k_{\|}\right)\right|^{2}\left(\frac{1}{\Delta r_{z}}\right).
\end{align}
The term $1/\Delta r_z$ is the normalization factor, in which $\Delta r_z$ is the length of sightline under consideration. To reveal the small-scale structures we are interested in, we select neutral patches with $\Delta r_z = 10$ comoving Mpc, and compute the 1D power spectra from segments of 10 comoving Mpc along the line of sight. For a reasonable number of $\mathcal{O}(10)$ high-$z$ background sources, the expected value of the power spectrum is obtained by averaging over 100 neutral patches on lines of sight
penetrating various environments,
i.e. $P\left(k_{\|}\right) \equiv\left\langle P\left(\hat{\boldsymbol{s}}, k_{\|}\right)\right\rangle$.
On each quasar spectrum, we will be able to select $\sim 10$ segments of 10 comoving Mpc length in neutral patches; as the neutral patches are intermittently separated by ionized regions during the EoR, we may need a spectrum covering $\sim 200$ comoving Mpc along the line of sight. A length of 200 comoving Mpc projects to a total bandwidth of about 14 MHz at redshift 9, corresponding to $\Delta z \sim 0.8$, which is reasonable in practice.
For the rest of the paper, we abbreviate $k_{\|}$ as $k$, as here we are always interested in the $k$-modes along the line of sight.

Supplementary Fig.~8 shows the evolution of the 1D power spectrum with redshift.
The solid lines in the left and middle panels show the power spectra in the CDM model and in the WDM model with $m_{\rm WDM} = 3$ keV respectively, in the absence of X-rays.
As the redshift increases, the halo abundance decreases, and the small-scale fluctuations in the forest signal decrease, resulting in steeper power spectra. The small-scale power is slightly more notably suppressed in the WDM model, as the halo formation is more delayed.
However, the redshift evolution has only a weak effect on the 1D power spectrum in the absence of X-ray heating.
The right panel of
Supplementary Fig.~8 illustrates the evolution of the 1D power spectrum in the CDM model with $f_{\rm X}$ = 3.
In the case of strong X-ray heating, the 1D power spectrum of the 21-cm forest is dramatically suppressed with the decreasing redshift, and the dominant reason is the rapidly increasing IGM temperature.
It implies that for the purpose of constraining DM properties, the 1D power spectrum measurement at higher redshift is preferred, as long as a radio-bright source at an even higher redshift is available.

\subsection{Measurement error on 1D power spectrum.}
\label{sub:psnoise}
The observational uncertainties in the 21-cm forest include the thermal noise, the sample variance, the contaminating spectral structures from foreground sources in the chromatic sidelobes, and the bandpass calibration error. The bandpass calibration error depends on specific calibration strategies, and mainly affects the broadband amplitude of the continuum, so we expect that it has a negligible effect on the small-scale features we are interested in. The contaminating spectral structures from foregrounds are not likely affecting the small structures we are aiming at, as the discriminating features locate at $k \gtrsim 3\, {\rm Mpc}^{-1}$, which are well within the ``EoR window''\citep{Thyagarajan2020}. Therefore, we consider only the thermal noise of an interferometer array, and the sample variance in the power spectrum measurement.

The sample variance on the 1D power spectrum is $P^S=\sigma_P(k)/\sqrt{N_s \times N_m}$, where $\sigma_P(k)$ is the standard deviation of $P(k)$ from $N_s\times N_m$ measurements of the 1D power spectrum at $k$, in which $N_s$ is the number of 1D power spectrum measurements on different neutral patches of $\Delta r_z$, and $N_m$ is the number of independent modes in each $k$-bin from each measurement.
Using 10 high-redshift background radio sources, it is reasonable to expect about 100 independent measurements of 1D power spectra from segments of spectra, each corresponding to a comoving length of 10 Mpc.
We adopt $N_s = 100$, and $\sigma_P(k)$ is obtained by simulating 21-cm forest signals from $N_s$ neutral segments of 10 comoving Mpc length penetrating various environments covering grid densities from $\delta = -0.7$ to $\delta = +1.5$.

As for the thermal noise error,
we follow the approach taken by Ref.~\cite{Thyagarajan2020}, and assume that each spectrum is measured for two times separately, or the total integration time is divided into two halves, and the cross-power spectrum is practically measured in order to avoid noise bias.
Then the observing time for each measurement of the spectrum is $\delta t_{0.5} = 0.5\, \delta t$, and the thermal noise on the spectrum is increased by a factor of $\sqrt{2}$.
Then the thermal noise uncertainty on the 1D power spectrum is given by \citep{Thyagarajan2020}
\begin{align}
P^{N} = \frac{1}{\sqrt{N_s}}\left(\frac{\lambda_{z}^{2} T_{{\rm sys}}}{ A_{{\rm eff}} \Omega}\right)^{2}
\left(\frac{\Delta r_{z}}{2 \Delta \nu_{z} \delta t_{0.5}}\right),
\end{align}
where $\Delta \nu_{z}$ is the total observing bandwidth corresponding to $\Delta r_{z}$.
A distance of 10 comoving Mpc along the line of sight corresponds to a bandwidth of $\Delta \nu_{z} = 0.56$ MHz at $z = 9$.
Assuming the same telescope parameters of SKA1-LOW and SKA2-LOW as those for the direct measurement, and the same observation time of $\delta t =$ 100 hr ($\delta t_{0.5}$ = 50 hr) on each source,
the expected thermal noise on the 1D power spectrum of 21-cm forest is plotted in Figs.~\ref{fig_PS_fw} and \ref{fig_PS_S150}, as well as in
Supplementary Fig.~8, with dotted lines for SKA1-LOW and dashed lines for SKA2-LOW, respectively.
The total measurement errors including the thermal noises of SKA2-LOW and sample variance are shown with the error bars in these figures.
We have tested the extraction of 21-cm forest 1D power spectrum by simulating mock quasar spectra with thermal noises, and calculating the 1D power spectra from the noisy spectra. The results are shown in
Supplementary Fig.~9, with upper panels from mock spectra with SKA1-LOW noises, and lower panels from mock spectra with SKA2-LOW noises, respectively. In each row, the left panel shows the results from mock spectra with both 21-cm absorption signals and thermal noises, and the right panel shows the results from mock spectra with only thermal noises. The measured noise power spectra agree well with the theoretical predictions.
It is seen that the measurement of 1D power spectrum notably improves the observability of the 21-cm forest signals as compared to the direct measurement of individual absorption lines.
With about 10 moderately bright quasars with $S_{\rm 150}\gtrsim 10$ mJy at redshift around 9,
the 1D power spectrum can be measured by SKA2-LOW even if the IGM was heated as sufficiently as in the model with $f_{\rm X} = 3$, and can reach a high signal-to-noise ratio if $f_{\rm X} \lesssim 1$. Note that the measurement error can be further suppressed if more sources are available beyond reionization, and more power spectra can be averaged to suppress both the thermal noise and the sample variance.




\clearpage
\begin{addendum}
\item[Data Availability]
The main data that support the results in this work are provided with this paper, and are also available at
\href{https://doi.org/10.57760/sciencedb.08093}{https://doi.org/10.57760/sciencedb.08093}.
Further datasets are available from the corresponding authors upon reasonable request.

\item[Code Availability]
The code {\tt 21cmFAST} used for large-scale simulation is publicly available at \\ \href{https://github.com/andreimesinger/21cmFAST}{https://github.com/andreimesinger/21cmFAST},
the codes for simulating small-scale structures and 21-cm forest signals are available from the corresponding authors upon reasonable request,
and the {\tt GADGET} code is available at \href{https://wwwmpa.mpa-garching.mpg.de/gadget}{https://wwwmpa.mpa-garching.mpg.de/gadget}.

\end{addendum}


\begin{addendum} 
\item[Additional information]

{\bf Correspondence and requests for materials} should be addressed to Yidong Xu (email: xuyd@nao.cas.cn), Xin Zhang (email: zhangxin@mail.neu.edu.cn) or Xuelei Chen (email: xuelei@cosmology.bao.ac.cn).

\item
We thank the anonymous referees for very constructive comments and suggestions.
We thank Yichao Li, Peng-Ju Wu, Jing-Zhao Qi, and Bin Yue for helpful discussions.
This work was supported by National Key R\&D Program of China (Grant No. 2022YFF0504300),
the National Natural Science Foundation of China (Grant Nos. 11973047, 11975072, 11835009, 11988101, and 12022306),
and the National SKA Program of China (Grant Nos. 2020SKA0110401, 2020SKA0110100, 2022SKA0110200, and 2022SKA0110203).
Y.X. and X.C. also acknowledge support by the CAS grant (Grant No. ZDKYYQ20200008).
Y.W. acknowledges support by the CAS Interdisciplinary Innovation Team (Grant No. JCTD-2019-05).
R.L. acknowledges support by the CAS grant (Grant No. YSBR-062) and
the grant from K.C.Wong Education Foundation.

\item[Author contributions]

Y.S. performed most of the computation and analysis, and wrote part of the manuscript. Y.X. led the study, contributed to the simulations, and wrote the majority of the manuscript. Y.W. and W.Y. contributed to the computation of the 1D power spectrum. Y.X. and R.L. proposed the study. X.Z. and X.C. contributed to the collaboration organization, the Fisher forecasts, and the manuscript writing, and supervised the study. All authors discussed the results and commented on the manuscript.

\item[Competing Interests]

The authors declare no competing interests.

\end{addendum}

\newpage

\newpage

\newpage
\begin{figure}
\centering
\includegraphics[angle=0, width=16.0cm]{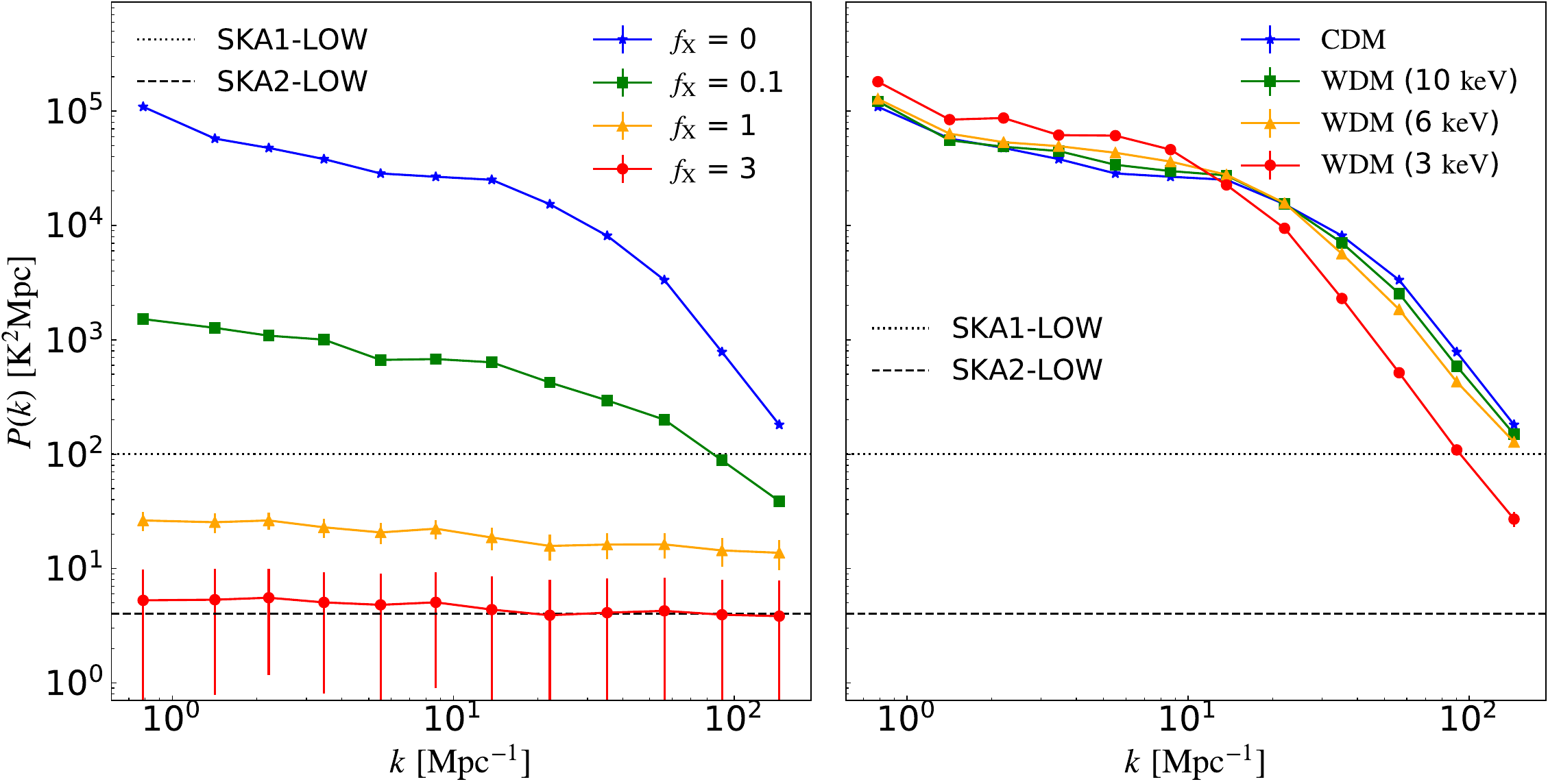}
\caption{\label{fig_PS_fw}
  \textbf{The expected 1D power spectrum of the 21-cm forest with $S_{150}$ = 10 mJy.}
  The expected 1D power spectrum of 21-cm forest at $z$ = 9 from a total of 100 measurements on segments of 10 comoving-megaparsec length in neutral patches along lines of sight against 10 background sources with $S_{\rm 150}$ = 10 mJy.
  The left panel shows the 1D power spectra in the CDM model, and the blue, green, yellow and red curves correspond to $f_{\rm X}$ = 0, 0.1, 1 and 3, respectively.
  The right panel shows the 1D power spectra for an un-heated IGM ($f_{\rm X} = 0$), and the blue, green, yellow and red curves correspond to the CDM model and the WDM models with $m_{\rm WDM}$ = 10 keV, 6 keV, and 3 keV, respectively.
  The black dotted and dashed lines in each panel are the thermal noises $P^{\rm N}$ expected for SKA1-LOW and SKA2-LOW respectively, with $\delta t$ = 100 hr, and the error bars show the total measurement errors of SKA2-LOW.
}
\end{figure}

\newpage
\begin{figure}
\centering
\includegraphics[angle=0, width=16.0cm]{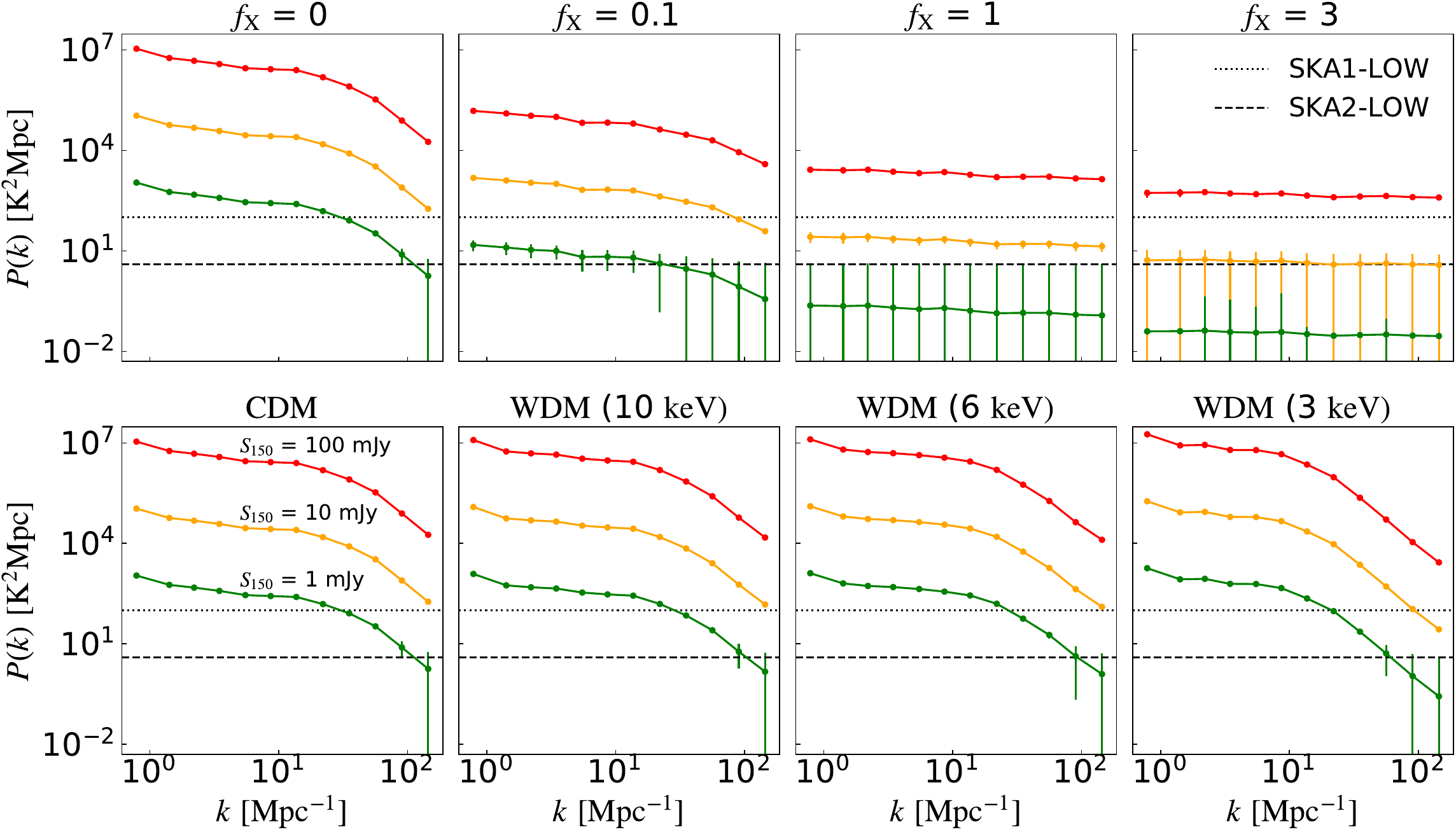}
\caption{\label{fig_PS_S150}
  \textbf{The expected 1D power spectrum of 21-cm forest for different heating histories and different DM models.}
  The expected 1D power spectrum of 21-cm forest at $z$ = 9 for different heating histories (top) and different DM models (bottom),
  assuming a total of 100 measurements on segments of 10 comoving-megaparsec length in neutral patches along lines of sight against 10 background sources.
  The upper panels show the power spectra in the CDM model assuming $f_{\rm X}$ = 0, 0.1, 1, and 3, from left to right respectively.
  The lower panels show the power spectra for the CDM model and the WDM models with $m_{\rm WDM}$ = 10 keV, 6 keV, and 3 keV, from left to right respectively, assuming an un-heated IGM ($f_{\rm X} = 0$).
  In each row, the green, yellow and red curves correspond to the flux densities of the background point sources with $S_{\rm 150}$ = 1 mJy, 10 mJy and 100 mJy, respectively.
  The black dotted and dashed lines are the thermal noises $P^{\rm N}$ for SKA1-LOW and SKA2-LOW respectively, with $\delta t$ = 100 hr, and the error bars show the total measurement errors of SKA2-LOW.
}
\end{figure}

\newpage
\begin{figure}
\centering
\includegraphics[angle=0, width=16.0cm]{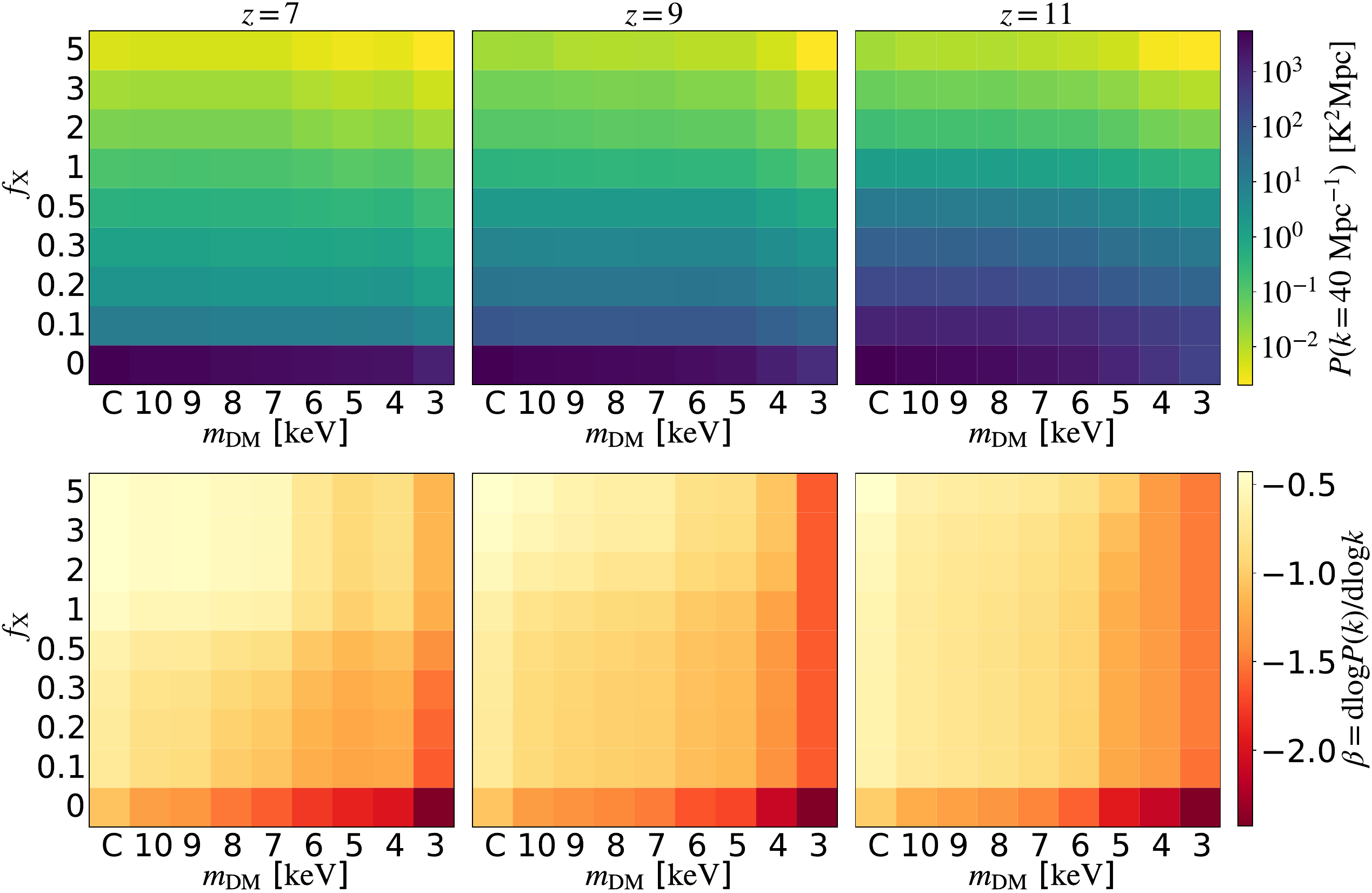}
\caption{\label{fig_magnitude_slope}
  \textbf{The amplitude and slope of the 1D power spectrum of the 21-cm forest.}
  The amplitude (top) and slope (bottom) of 1D power spectrum of 21-cm forest
  at $k$ = 40 ${\rm Mpc}^{-1}$.
  The abscissa and ordinate represent different $m_{\rm DM}$ and $f_{\rm X}$, respectively, and ``C'' stands for the CDM model.
  From left to right, the redshift of each panel is 7, 9 and 11.
  The flux density of the background source is assumed to be $S_{150}=10$ mJy.
  }
\end{figure}

\newpage
\begin{figure}
\centering
\includegraphics[angle=0, width=16.0cm]{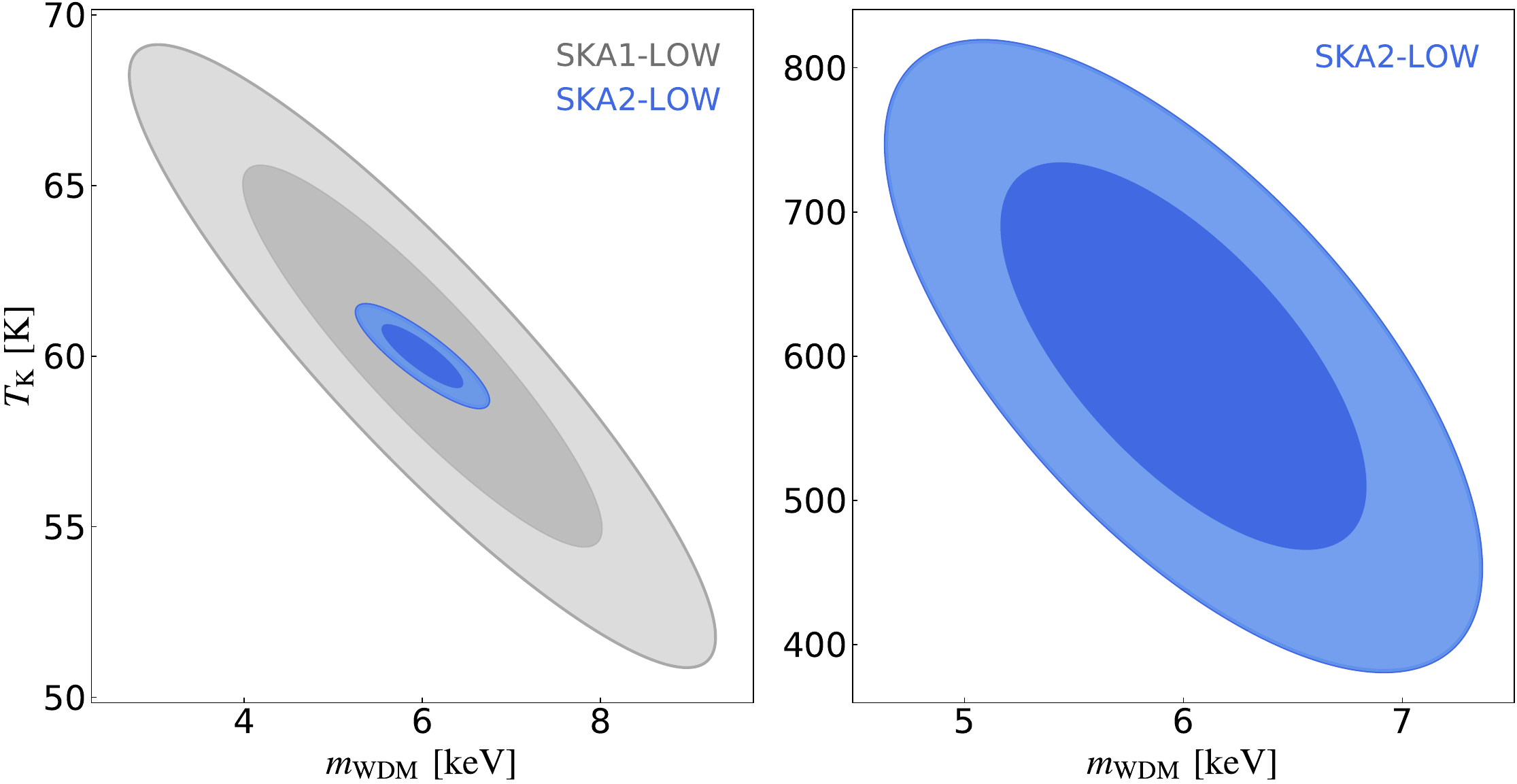}
\caption{\label{fig_contour}
 \textbf{Constraints on $T_{\rm K}$ and $m_{\rm WDM}$ with the 1D power spectrum of the 21-cm forest with $S_{150}$ = 10 mJy.}
 Constraints (68.3$\%$ and 95.4$\%$ confidence level) on $T_{\mathrm{K}}$ and $m_{\mathrm{WDM}}$ with the 1D power spectrum of 21-cm forest at $z$ = 9,
assuming a total of 100 measurements on segments of 10 comoving-megaparsec length in neutral patches along lines of sight against 10 background sources
with $S_{150}$ = 10 mJy.
The gray and blue contours correspond to results for SKA1-LOW and SKA2-LOW, respectively, including the sample variance and the thermal noise with observation of 100 hr on each source.
 The fiducial model of the left panel is $m_{\mathrm{WDM}}$ = 6 keV and $T_{\mathrm{K}}$ = 60 K (corresponding to $f_{\mathrm{X}}$ = 0.1),
 and the fiducial model of the right panel is $m_{\mathrm{WDM}}$ = 6 keV and $T_{\mathrm{K}}$ = 600 K (corresponding to $f_{\mathrm{X}}$ = 1).
  }
\end{figure}

\bibliography{21cmref}



\newpage
\begin{efigure}
\centering
\includegraphics[angle=0, width=16.0cm]{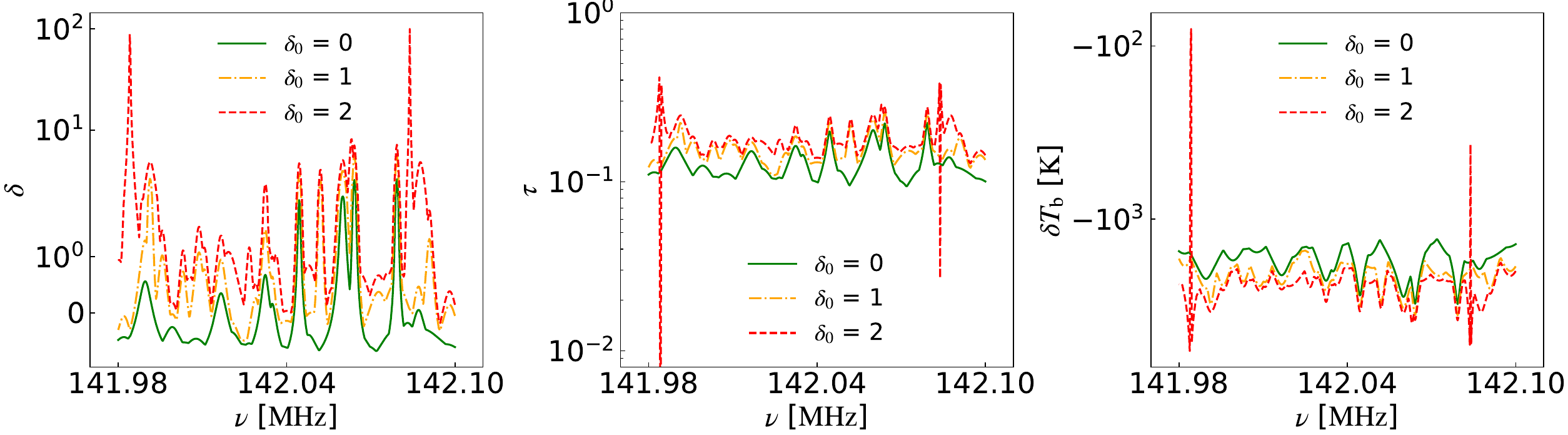}
\caption{\label{fig_delta}
  \textbf{The density (left panel), optical depth (middle panel) and brightness temperature (right panel) for a line of sight of 2 comoving Mpc in the CDM model at $z$ = 9.}
  The green, yellow and red lines correspond to local overdensities of $\delta_0$ = 0, 1 and 2, respectively.
  The flux density of the background source in the right panel is assumed to be $S_{150}=10$ mJy. }
\end{efigure}

\begin{efigure}
\centering
\includegraphics[angle=0, width=8.0cm]{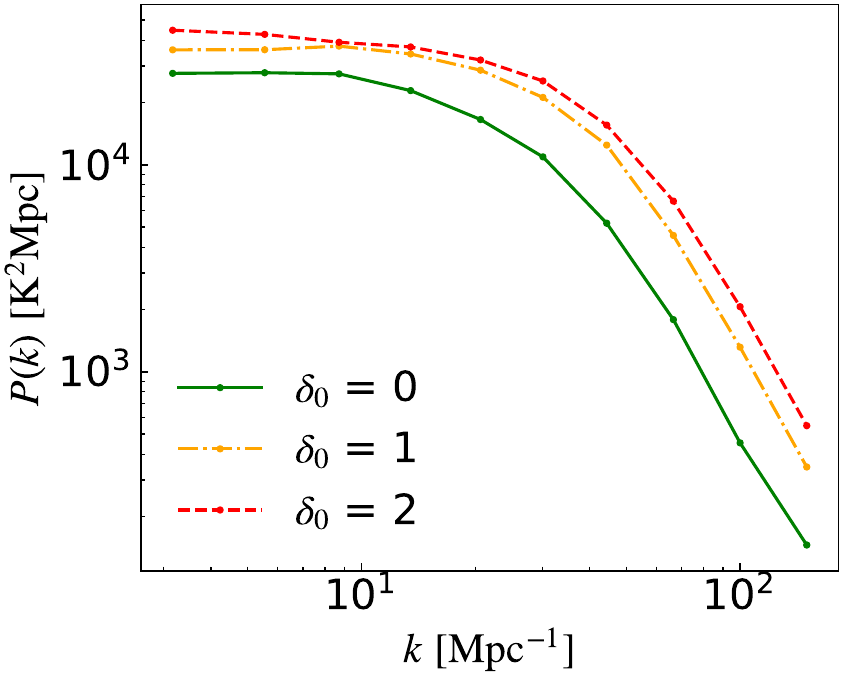}
\caption{\label{fig_delta_PS}
  \textbf{1-D power spectrum of a synthetic 21-cm forest spectrum in the CDM model, for a line of sight penetrating through an un-heated IGM ($f_{\rm X} = 0$) with different local overdensities at $z$ = 9.}
  The green, yellow and red curves correspond to $\delta_0$ = 0, 1 and 2, respectively.
  The flux density of the background source is assumed to be $S_{150}=10$ mJy.
  }
\end{efigure}

\newpage
\begin{sfigure}
\centering
\includegraphics[angle=0, width=10.0cm]{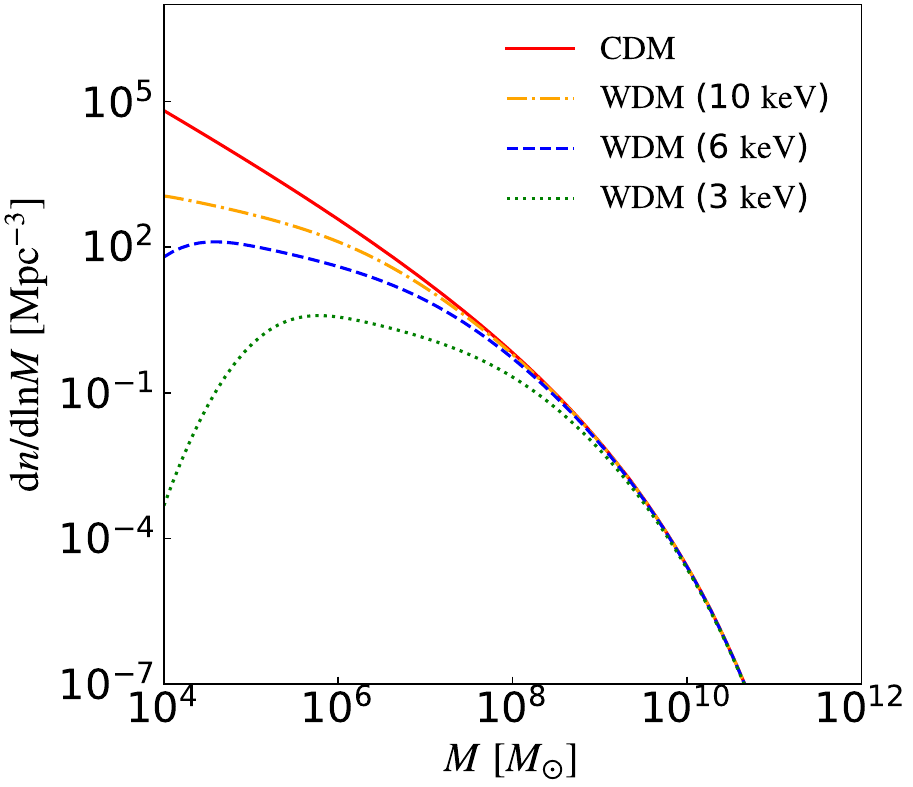}
\caption[\textbf{Supplementary Figure \arabic{figure}}]{\label{fig_hmf}
  \textbf{Halo mass function for different DM particle masses at $z$ = 9.}
The red, yellow, blue and pink curves correspond to the CDM model and WDM models with $m_{\rm WDM}$ = 10 keV, 6 keV, and 3 keV, respectively.
  }
\end{sfigure}

\newpage
\begin{sfigure}
\centering
\includegraphics[angle=0, width=10.0cm]{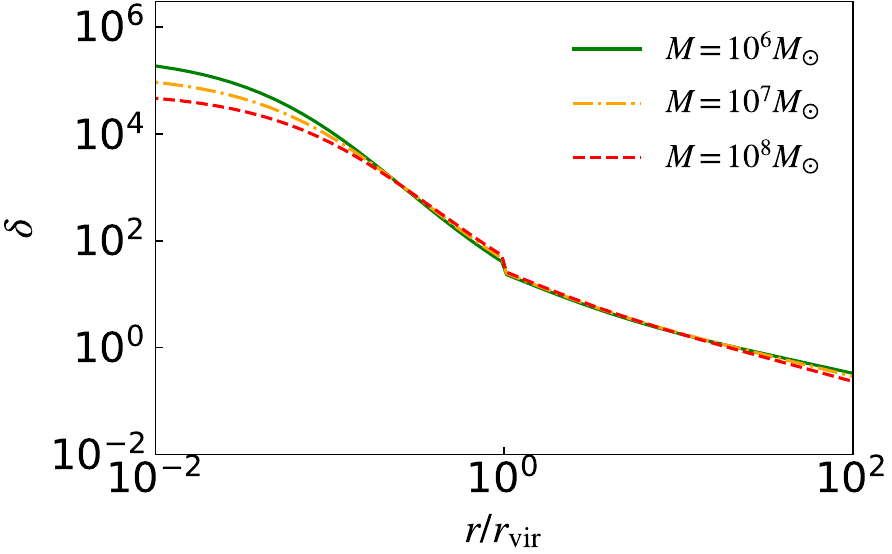}
\caption[\textbf{Supplementary Figure \arabic{figure}}]{\label{fig_pro_delta}
  \textbf{Neutral hydrogen overdensity profiles inside and outside the virial radius of a halo at $z$ = 9.}
  The green, yellow and red lines correspond to halo mass of $10^6  M_{\odot}$, $10^7  M_{\odot}$ and $10^8  M_{\odot}$, respectively.
  }
\end{sfigure}

\begin{sfigure}
\centering
\includegraphics[angle=0, width=10.0cm]{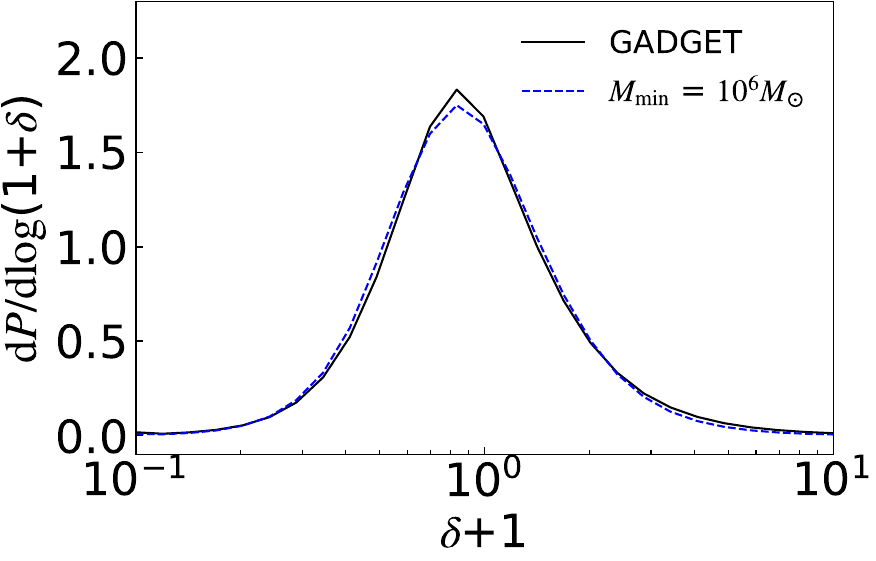}
\caption[\textbf{Supplementary Figure \arabic{figure}}]{\label{fig_pdf}
  \textbf{Probability density distribution of the gas overdensity at $z$ = 17.}
The black solid line is the probability density distribution from the {\tt GADGET} simulation with a box size of 4 $h^{-1}{\rm Mpc}$ and $2\times800^3$ gas and DM particles.
The blue dashed line is the one derived from our hybrid approach with the same resolution as the {\tt GADGET} simulation.
  }
\end{sfigure}

\newpage
\begin{sfigure}
\centering
\includegraphics[angle=0, width=16.0cm]{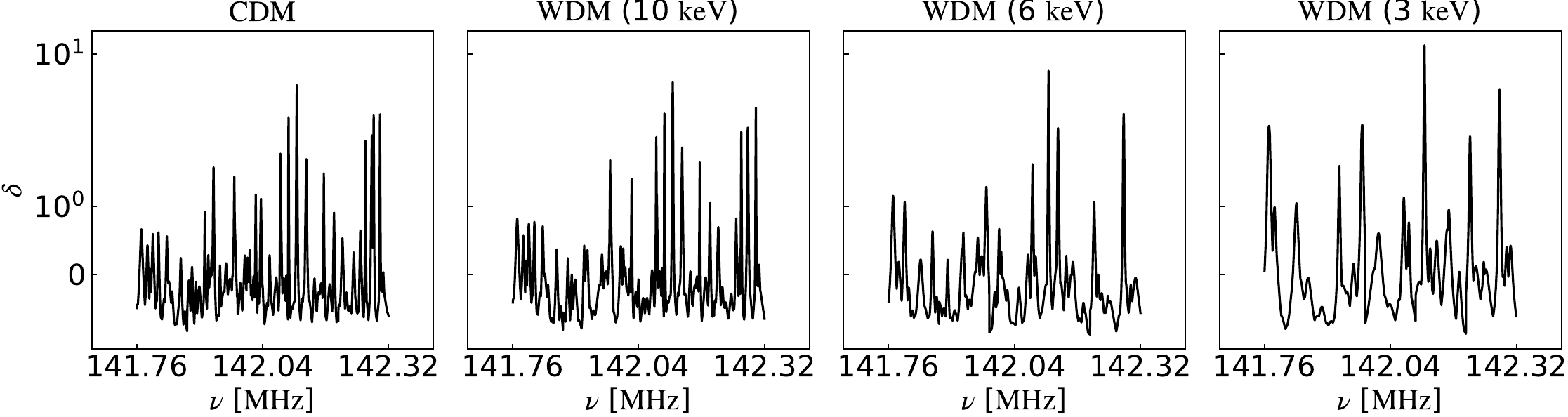}
\caption[\textbf{Supplementary Figure \arabic{figure}}]{\label{fig_density}
  \textbf{Density distribution of a patch of 10 comoving Mpc at $z$ = 9 along the line of sight, for an un-heated IGM ($f_{\rm X}$ = 0).}
The four panels, from left to right, correspond to the CDM model and the WDM models with $m_{\rm WDM}$ = 10 keV, 6 keV and 3 keV, respectively.
  }
\end{sfigure}

\begin{sfigure}
\centering
\includegraphics[angle=0, width=10.0cm]{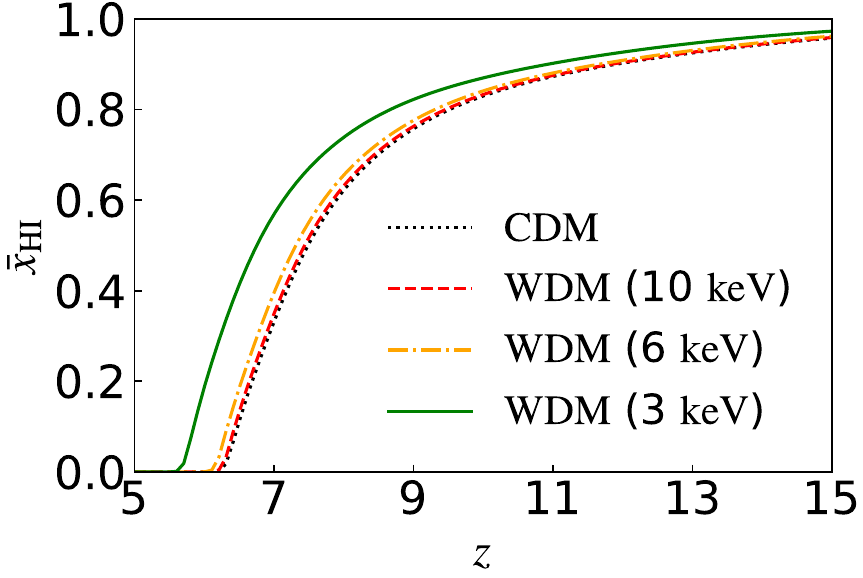}
\caption[\textbf{Supplementary Figure \arabic{figure}}]{\label{fig_xi_history}
  \textbf{Reionization history simulated by {\tt 21cmFAST.}}
  The black, red, yellow and green curves correspond to the average neutral fraction $\bar{x}_{\rm HI}$ as a function of redshift $z$ in the CDM model and the WDM models with $m_{\rm WDM}$ = 10 keV, 6 keV and 3 keV, respectively.
  }
\end{sfigure}

\newpage
\begin{sfigure}
\centering
\includegraphics[angle=0, width=10.0cm]{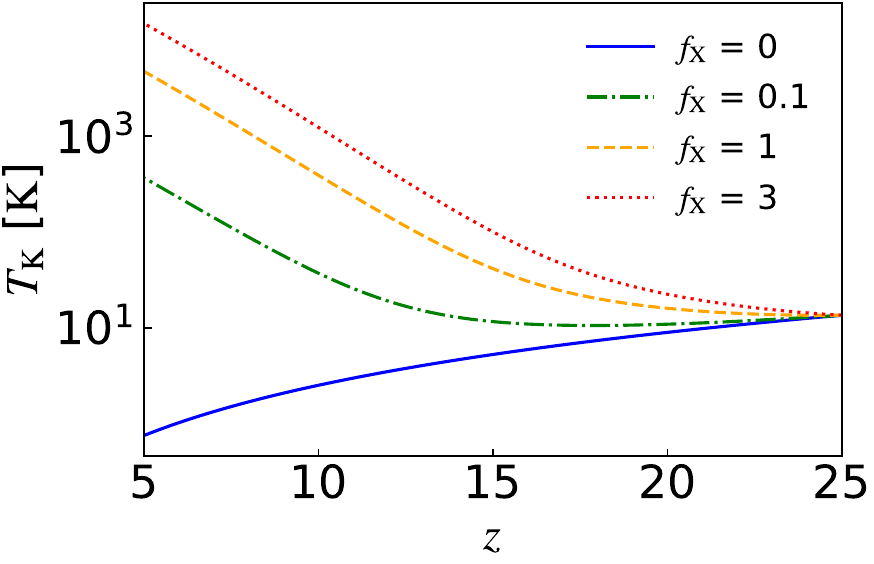}
\caption[\textbf{Supplementary Figure \arabic{figure}}]{\label{fig_Tk}
  \textbf{Evolution of the global gas temperature with redshift.}
  The blue, green, yellow and red lines correspond to $f_{\rm X}$ = 0, 0.1, 1 and 3, respectively.
  }
\end{sfigure}

\begin{sfigure}
\centering
\includegraphics[angle=0, width=10.0cm]{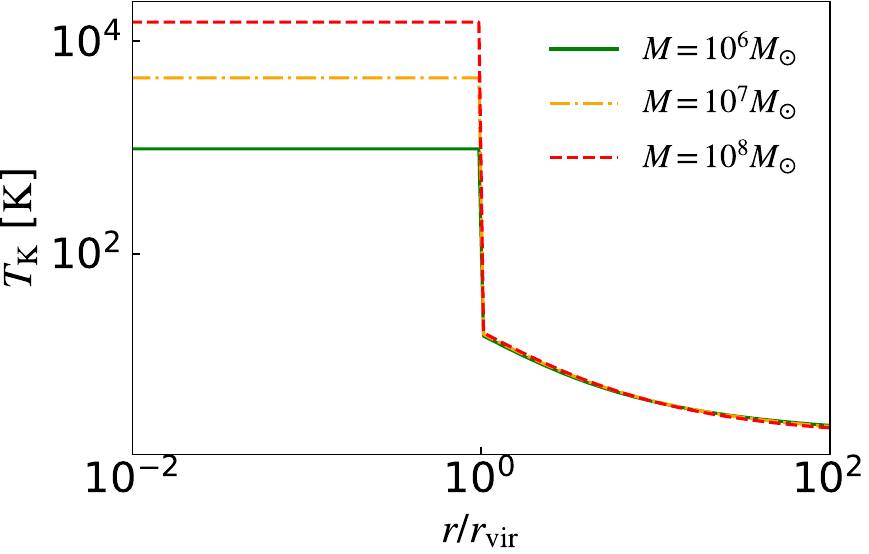}
\caption[\textbf{Supplementary Figure \arabic{figure}}]{\label{fig_pro_Tk}
  \textbf{Temperature profiles of gas inside and outside the virial radii of halos at $z$ = 9 with an un-heated IGM ($f_{\rm X}$ = 0).}
  The green, yellow and red lines correspond to halo masses of $10^6  M_{\odot}$, $10^7  M_{\odot}$ and $10^8  M_{\odot}$, respectively.
  }
\end{sfigure}

\newpage
\begin{sfigure}
  \centering
  \includegraphics[angle=0, width=16.0cm]{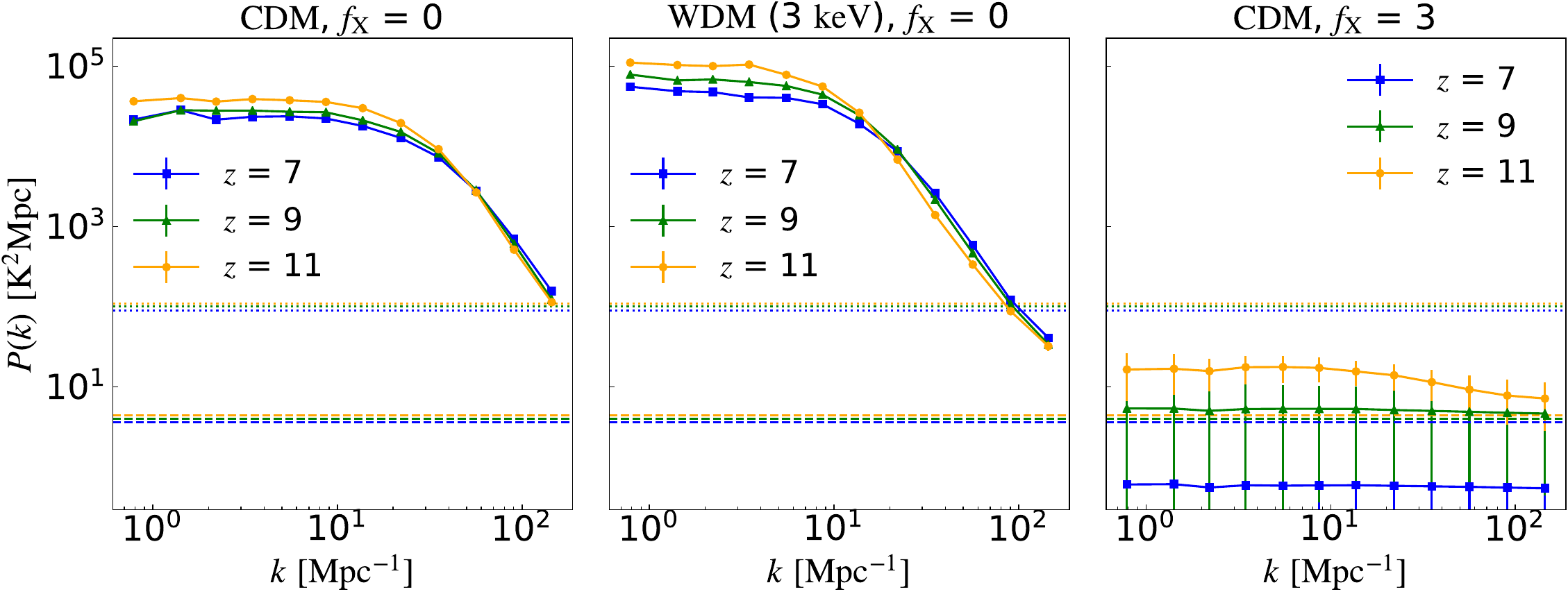}
  \caption[\textbf{Supplementary Figure \arabic{figure}}]{\label{fig_PS_z}
  \textbf{Evolution of the 1-D power spectrum of 21-cm forest averaged over 100 measurements on segments of 10 comoving Mpc length in neutral patches along lines of sight against background sources
with $S_{150}$ = 10 mJy.}
  The solid lines in the left and central panels show the power spectra in the CDM model and those in the WDM model with $m_{\rm WDM}$ = 3 keV respectively, assuming an un-heated IGM ($f_{\rm X}$= 0).
  The solid lines in the right panel show the power spectra in the CDM model assuming an efficiently-heated IGM ($f_{\rm X}$ = 3).
  In each panel, the blue, green and yellow lines correspond to $z$ = 7, 9 and 11, respectively.
  The dotted and dashed lines with the corresponding colors are the expected thermal noises $P^{\rm N}$ for SKA1-LOW and SKA2-LOW, respectively, and the error bars show the total measurement errors of SKA2-LOW.
  }
\end{sfigure}

\newpage
\begin{sfigure}
  \centering
  \includegraphics[angle=0, width=16.0cm]{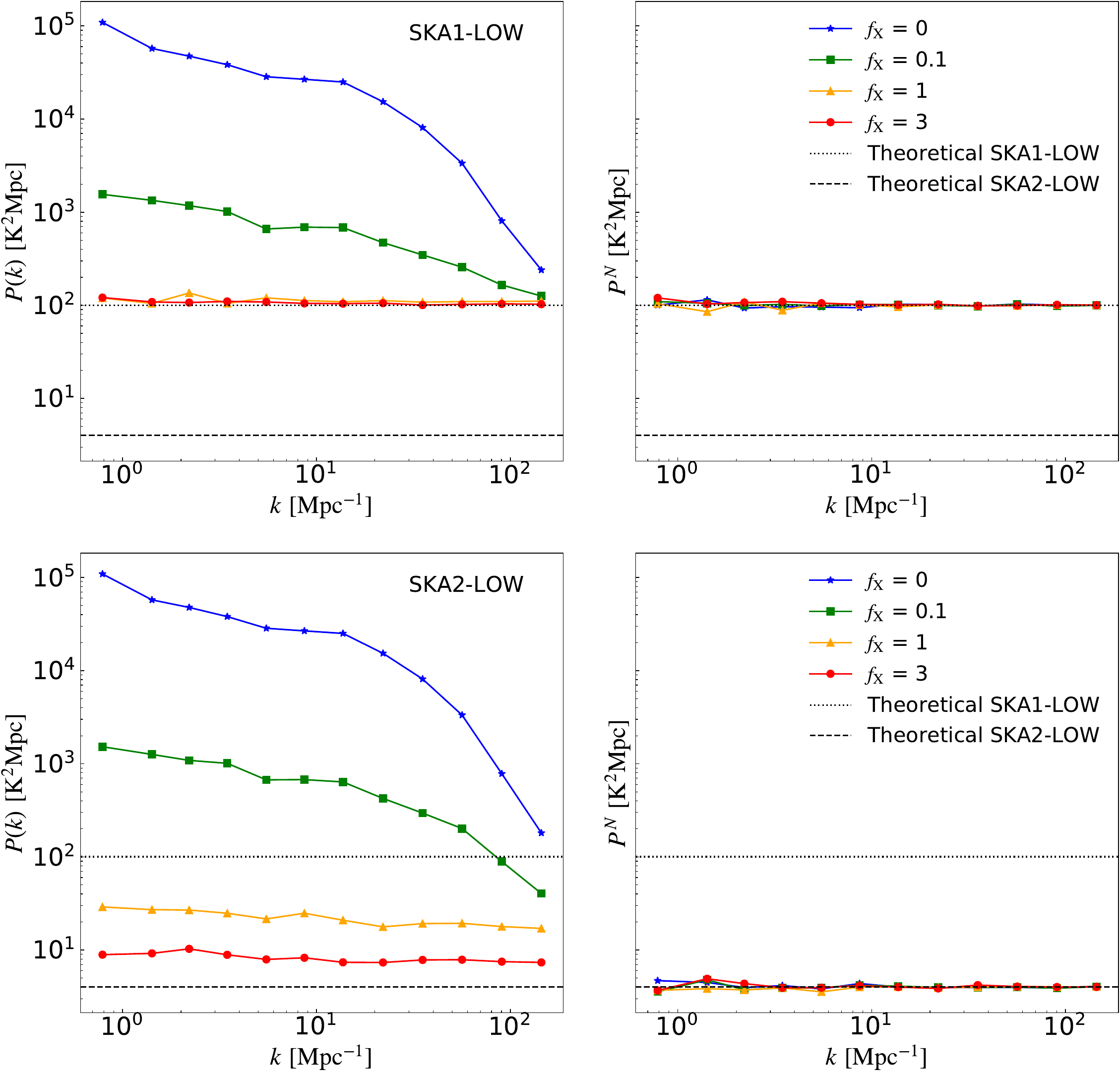}
\caption[\textbf{Supplementary Figure \arabic{figure}}]{
\textbf{1-D cross-power spectrum computed from mock spectra simulated with thermal noises expected for SKA1-LOW (upper panels) and SKA2-LOW (lower panels), respectively.}
The left plots show the results in which the mock spectra contain both 21-cm forest signal and thermal noise, and the right plots show the results from
mock spectra with only thermal noise.
Same as Fig.~3
the 1-D power spectra are averaged over 100 measurements on segments of 10 comoving Mpc length in neutral patches along lines of sight against 10 background sources
with $S_{150}$ = 10 mJy.
The blue, green, yellow and red curves correspond to $f_{\rm X}$ = 0, 0.1, 1 and 3, respectively.
The dotted and dashed lines are the theoretical thermal noises $P^N$ expected for the SKA1-LOW and SKA2-LOW, respectively.
\label{sfig_mockPS}}
\end{sfigure}

\end{document}